\documentclass[onecolumn]{IEEEtran}

\usepackage{symbols}
\usepackage{makeidx}  
\usepackage{latexsym}
\usepackage{amsmath, amssymb, ascmac, mathrsfs}
\usepackage[dvipdfm]{graphicx}
\usepackage{amsfonts, url}
\usepackage{float}
\usepackage{color}

\newtheorem{teigi}{Definition}
\newtheorem{teiri}{Theorem}      
\newtheorem{hodai}{Lemma}  
\newtheorem{tyuui}{Remark}
\newtheorem{kei}{Corollary}
\newtheorem{meidai}{Proposition}
\newtheorem{claim}{Claim}

\renewcommand{\c}[1]{{\cal #1}}

\newcommand{\bb}[1]{{\mathbb #1}}

\makeatletter
\def\eqnarray{%
        \stepcounter{equation}%
        \let\@currentlabel=\theequation
        \global\@eqnswtrue\global\@eqcnt\z@
        \tabskip\@centering
        \let\\=\@eqncr
        $$\halign to \displaywidth\bgroup\@eqnsel\hskip\@centering
        $\displaystyle\tabskip\z@{##}$&\global\@eqcnt\@ne
        \hfil$\displaystyle{{}##{}}$\hfil
        &\global\@eqcnt\tw@$\displaystyle\tabskip\z@{##}$\hfil
        \tabskip\@centering&\llap{##}\tabskip\z@\cr}
\makeatother

\makeatletter

\def\QED{{\unskip\nobreak\hfil\penalty50
\hskip1em\hbox{}\nobreak\hfil \hbox{$\square$\hskip1em}
\parfillskip\z@ \finalhyphendemerits\z@\par}}

\allowdisplaybreaks

\def\QED{{\unskip\nobreak\hfil\penalty50
\hskip1em\hbox{}\nobreak\hfil \hbox{$\square$\hskip1em}
\parfillskip\z@ \finalhyphendemerits\z@\par}}

\allowdisplaybreaks

\begin{document}

\title{Security Formalizations and Their Relationships for Encryption and Key Agreement in Information-Theoretic Cryptography}

\author{Mitsugu Iwamoto,~\IEEEmembership{Member,~IEEE,}
 Kazuo Ohta,~\IEEEmembership{Member,~IEEE,}
 and~Junji~Shikata,~\IEEEmembership{Member,~IEEE,}
\thanks{This paper was presented in part at 2011 IEEE International Symposium on Information Theory (ISIT2011) \cite{IO11} and 2013 IEEE International Symposium on Information Theory (ISIT2013) \cite{Shikata13}.}
\thanks{M. Iwamoto is with Center for Frontier Science and Engineering, the University of Electro-Communications, Japan. K. Ohta is with Graduate School of Informatics and Engineering, the University of Electro-Communications, Japan. J. Shikata is with Graduate School of Environment and Information Sciences, Yokohama National University, Japan. 
(e-mail: mitsugu@inf.uec.ac.jp, ota@inf.uec.ac.jp, shikata@ynu.ac.jp)}
}

\maketitle

\begin{abstract}
This paper revisits formalizations of information-theoretic security for symmetric-key encryption and key agreement protocols which are very fundamental primitives in cryptography.  In general, we can formalize information-theoretic security in various ways: some of them can be formalized as stand-alone security by extending (or relaxing) Shannon's perfect secrecy or by other ways such as semantic security; some of them can be done based on composable security. Then, a natural question about this is: what is the gap between the formalizations? To answer the question, we investigate relationships between several formalizations of information-theoretic security for symmetric-key encryption and key agreement protocols. Specifically, for symmetric-key encryption protocols in a general setting including the case where there exist decryption-errors, we deal with the following formalizations of security: formalizations extended (or relaxed) from Shannon's perfect secrecy by using mutual information and statistical distance; information-theoretic analogues of indistinguishability and semantic security by Goldwasser and Micali; 
and composable security by Maurer et al. and Canetti.   
Then, we explicitly show the equivalence and non-equivalence between those formalizations. Under the model, we also derive lower bounds on the adversary's (or distinguisher's) advantage and the size of secret-keys required under all of the above formalizations. Although some of them may be already known, we can explicitly derive them all at once through our relationships between the formalizations. In addition, we briefly observe impossibility results which easily follow from the lower bounds. The similar results are also shown for key agreement protocols in a general setting including the case where  there exist agreement-errors in the protocols. 
\end{abstract}


\IEEEpeerreviewmaketitle

\section{Introduction}
{\bf Background and Related Works.} 
The security of cryptographic protocols in information-theoretic cryptography does not require any computational assumption based on computationally hard problems, such as the integer factoring and discrete logarithm problems. In addition, since the security definition in information-theoretic cryptography is formalized by use of some information-theoretic measure (e.g. entropy or statistical distance) or some probability (e.g., success probability of adversary's guessing), it does not depend on a specific computational model and can provide security which does not compromise even if computational technology intensively develops or a new computational technology (e.g. quantum computation) appears in the future. In this sense, it is interesting to study and develop cryptographic protocols with information-theoretic security.        

As fundamental cryptographic protocols we can consider symmetric-key encryption and key-agreement protocols, and the model of the protocols falls into a very simple and basic scenario where there are two honest players (named Alice and Bob) and an adversary (named Eve). 
Up to date, various results on the topic of those protocols with information-theoretic security have been reported and developed since Shannon's work \cite{Shannon}. In most of those results the traditional security definition has been given as {\it stand-alone security} in the sense that the protocols will be used in a stand-alone way: in symmetric-key encryption, the security is formalized as $I(M;C)=0$ (Shannon's perfect secrecy) or its variant (e.g. $I(M;C)\le \epsilon$ for some small $\epsilon$), where $M$ and $C$ are random variables which take values in sets of plaintexts and ciphertexts, respectively; similarly, in key agreement the security is usually formalized as $I(K;T)=0$ or its variant (e.g. $I(K;T)\le \epsilon$), where $K$ and $T$ are random variables which take values on sets of shared keys and transcripts, respectively. 
In addition, it is possible to give security formalizations of symmetric-key encryption by an information-theoretic analogue of indistinguishability or semantic security by Goldwasser and Micali \cite{GM}. 
The problem with those definition of stand-alone security is that, if a protocol is composed with other ones, the security of the combined protocol may not be clear. Namely, it is not always guaranteed that the composition of individually {\it secure} protocols results in the {\it secure} protocol, where {\it secure} is meant in the sense of the traditional definition of stand-alone security. 

On the other hand, {\it composable security} (or security under composition) can guarantee that a protocol remains to be secure after composed with other ones. The previous frameworks of this line of researches are based on the {\it ideal-world/real world paradigm}, and the paradigm includes {\it universal composability} by Canetti \cite{C05} and {\it reactive simulatability} by Backes, Pfitzmann and Waidner \cite{BPW} (See also \cite{C01,PW01,GL,MR,Beaver} for related works). In addition, the explicit and simple paradigm for composable  security was given by Maurer \cite{Maurer10}, and this approach is called {\it constructive cryptography} where the security definitions of cryptographic systems can be understood as constructive statements: the idea is to consider cryptographic protocols as transformations which construct cryptographically {\it stronger} systems from {\it weaker} ones. 
Using the framework of constructive cryptography, Maurer and Tackmann \cite{MT10} studied the authenticate-then-encrypt paradigm  for symmetric-key encryption with computational security. 
Furthermore, Maurer and Renner \cite{MR11} proposed a new framework in an abstract way, called {\it abstract cryptography}. The framework is described at a higher level of abstraction than \cite{Maurer10,MT10}, and various notions and methodologies (e.g. universal composability \cite{C05}, reactive simulatability \cite{BPW}, and indifferentiability \cite{MRH}) can be captured in the framework. 

Up to date, there are a few works which report a gap between formalizations of the stand-alone security and composable security for multiparty computation in information-theoretic settings \cite{BMU07, DM00, KLR06}. In particular, Kushilevitz, Lindell and Rabin \cite{KLR06} investigated the gap between them in several settings (i.e., perfect/statistical security and composition with adaptive/fixed inputs), and they showed a condition that a protocol having stand-alone security is not necessarily secure under universal composition.     

\medskip    

\noindent
{\bf Our Contributions.}
We can formalize information-theoretic security for symmetric-key encryption and key agreement protocols in various ways: some of them can be formalized as stand-alone security by extending Shannon's perfect secrecy or by other ways such as semantic security; some of them can be done based on composable security. Then, a natural question about this is: what is the gap between the formalizations?  
To answer the question, we investigate relationships between several formalizations of information-theoretic security for symmetric-key encryption and key agreement protocols.  
Specifically, we deal with the model of symmetric-key encryption protocols in a general setting where encryption/decryption algorithms can be arbitrary (i.e., deterministic or randomized), or protocols can have decryption-errors. In the model, we investigate the following formalizations of security: 
\begin{enumerate}
\item[(i)] Traditional formalization extended (or relaxed) from Shannon's perfect secrecy by using the mutual information; 
\item[(ii)] Another traditional one extended (or relaxed) from Shannon's perfect secrecy by using the statistical distance (a.k.a. the variational distance); 
\item[(iii)] Formalization by information-theoretic analogue of indistinguishability by Goldwasser and Micali \cite{GM}; 
\item[(iv)] Formalization by information-theoretic analogue of semantic security by Goldwasser and Micali \cite{GM}; 
\item[(v)] Formalizations of composable security by Maurer et al. \cite{MR11, MT10} and Canetti \cite{C01,C05}.  
\end{enumerate} 

The main contribution of this paper is to explicitly show that relationships between those formalizations, and to reveal the conditions that those formalizations being (non)equivalent in details. 
Under the model, we also derive lower bounds on the adversary's (or distinguisher's) advantage and secret-key size required under all of the above formalizations. Although some of them may be already known, we can explicitly derive them all at once through our relationships between the formalizations in combination with the lower bound shown by Pope \cite{POPE08} in which  the security definition is given based on Maurer's formalization for composable security.     
In addition, we briefly observe impossibility results which easily follow from the lower bounds. 

Furthermore, we show similar results (i.e., relationships between formalizations, lower bounds, and impossibility results) for key agreement protocols in a general setting where the channels used are unidirectional/bidirectional, the round number of protocols is arbitrary, and the protocols can have agreement-errors. 

\medskip
\noindent
{\bf Other Works Related.} 
Bellare, Tessaro, and Vardy \cite{BTV12} recently study security definitions and schemes for encryption in the model of the wiretap channels \cite{Wyner}. In particular, in the model of wiretap channels, they showed that the following formalizations of stand-alone security are equivalent: formalizations extended (or relaxed) from Shannon's perfect secrecy by using the mutual information and statistical distance; information-theoretic indistinguishability which is called {\it distinguishing security} in \cite{BTV12}; and information-theoretic semantic security. 
Although the main scope of their paper lies in the wiretap channel and it is different from the model in this paper, their approach and ours are similar. 
They also showed that the first formalization by using mutual information with restriction on that only uniformly distributed plaintexts are input is weaker than those formalizations.   

Recently, in a simple and elemental way, Dodis \cite{Dodis12} directly derives a lower bound on secret-key size required for symmetric-key encryption, which may have decryption-errors, with specifying required running time of an adversary where the security definition is given based on a simulation-based formalization under bounded/unbounded adversaries. One of lower bounds in Corollary \ref{bound104} in this paper is the same as his lower bound, and interestingly, our technique and his for deriving it are quite different.     

\medskip

\noindent
{\bf Organization.} 
The rest of this paper is organized as follows. In Section II, we survey composable security and its formalization based on \cite{MR11, MT10} which is similar in spirit to previous ones in \cite{BPW, C01,C05,PW01}. 
In Section III, we explain the protocol execution of symmetric-key encryption in a general setting, and we give several formalizations of correctness and security in our model. 
Section IV is devoted to the main contribution of the paper, and we show the relationships between those formalizations for symmetric-key encryption protocols, and reveal conditions for equivalence and non-equivalence of the formalizations. In addition, we derive lower bounds on adversary's (or distinguisher's) advantage and the size of secret-keys required under all the formalizations. Furthermore, impossibility results are briefly observed. In Section V, we show similar results for key agreement protocols as well. 
Finally, we conclude the paper in Section VI.  

\medskip

\noindent
{\bf Notation.} In this paper, for a random variable $X$ which takes values in a finite set ${\cal X}$, the min-entropy and Hartley entropy of $X$ (i.e., log of the cardinality of the set) are denoted by $H_{\infty }(X)$ and $H_0(X)$, respectively. Also, $I(X;Y)$ denotes the mutual information between $X$ and $Y$, and we denote the statistical distance between two distributions $P_{X}$ and $P_{Y}$ by $\Delta(P_X,P_Y)$. For completeness, we describe the definitions in Appendix A. 

For an $n$-tuples of random variables $(X_1,X_2,\ldots,X_n)$, we denote its associated probability distribution by $P_{X_1 X_2 \ldots X_n}$. In this paper, for a random variable $X$ which takes values in ${\cal X}$, we especially write $P_{XX}$  for the distribution on ${\cal X} \times {\cal X}$ defined by $P_{XX}(x,x'):=P_X(x)$ if $x=x'$, and $P_{XX}(x,x'):=0$ if $x\not=x'$. 
Furthermore,  $|{\cal X}|$ denotes the cardinality of ${\cal X}$. Also, let ${\cal P}({\cal X})$ be the set of all distributions over ${\cal X}$ whose supports are ${\cal X}$, i.e., ${\cal P}({\cal X}):=\{ P_X \mid \mbox{Supp}(P_X)={\cal X} \}$.

\section{Composable Security}\label{SS}
In this paper, we consider a very basic scenario where there are three entities, Alice, Bob (honest players), and Eve (an adversary). 

\subsection{Definition of Systems}  
Following the notions in \cite{MR11,MT10}, we describe three types of systems: resources, converters and distinguishers (See \cite{MR11,MT10} for more details).  

A {\it resource} is a system with three interfaces labeled $A$, $B$, and $E$, where $A$, $B$, and $E$ imply three entities, Alice, Bob, and Eve, respectively. If two resources $R,S$ are used in parallel, this system is called {parallel composition} of $R$ and $S$ and denoted by $R \parallel S$. We note that $R \parallel S$ is also a resource. 

A {\it converter} is a system with two kinds of interfaces: the first kind of interfaces are designated as the {\it inner} interfaces which can be connected to interfaces of a resource, and combining a converter and a resource by the connection results in a new resource; the second  kind of interfaces are designed as the {\it outer} interfaces which can be provided as the new interfaces of the combined resource.  
For a resource $R$ and a converter $\pi$, we write $\pi (R)$ for the system obtained by combining $R$ and $\pi$, and $\pi (R)$ behaves as a resource, again. A {\it protocol} is a pair of converters $\pi=(\pi_A, \pi_B)$ for the honest players, Alice and Bob, and the resulting system by applying $\pi$ to a resource $R$ is denoted by $\pi (R)$  or $\pi_A \pi_B (R)$. 
For converters (or protocols) $\pi,\phi$, the {\it sequential composition} of them, denoted by $\phi \circ \pi$, is defined by $(\phi \circ \pi) (R):=\phi (\pi (R))$ for a resource $R$. In contrast, the {\it parallel composition} of converters (or protocols) $\pi,\phi$, denoted by $\pi \parallel \phi$, is defined by $(\pi \parallel \phi) (R \parallel S):=\pi(R) \parallel \phi(S)$ for resources $R,S$.      

A {\it distinguisher} for an $n$-interface resource is a system with $n+1$ interfaces: $n$ interfaces are connected to $n$ interfaces of the resource, respectively; and the other interface outputs a bit (i.e., 1 or 0). For a resource $R$ and a distinguisher $D$, we write {\it DR} for the system obtained by combining $R$ and $D$, and we regard {\it DR} as a binary random variable. 
The purpose of distinguishers is to distinguish two resources, and the advantage of a distinguisher $D$ for two resources $R_0, R_1$ is defined by    
\begin{eqnarray*}
\Delta^{D}(R_0,R_1):=\Delta(P_{DR_0}, P_{DR_1}),  
\end{eqnarray*}
where $P_{DR_0}$ and $P_{DR_1}$ are the probability distributions of the binary random variables $DR_0$ and $DR_1$, respectively.  
Let ${\cal D}$ be the set of all distinguishers, and we define 
\begin{eqnarray*}
\Delta^{{\cal D}}(R_0,R_1):=\sup_{D\in {\cal D}} \Delta^{D}(R_0,R_1).
\end{eqnarray*} 
Note that ${\cal D}$ contains not only polynomial-time distinguishers but also computationally unbounded ones, since this paper deals with information-theoretic security.      

\subsection{Definition of Security}\label{our_security}
The security definition we focus on in this paper is derived from the paradigm of constructive cryptography \cite{Maurer10}. Technically, the formal definition is based on the works in \cite{MR11, MT10} (see \cite{MR11, MT10} for details), and is similar in spirit to previous simulation-based definitions in \cite{BPW, C01,C05,PW01}. 
The idea in the paradigm of constructive cryptography includes comparison of the {\it real} and {\it ideal} systems: the real system means construction $\pi (R)$ by applying a protocol $\pi$ to a resource $R$; and the ideal system consists of the {\it ideal functionality} (such as ideal channels) $S$ including description of a security goal and a simulator $\sigma$ connected to the interface of $E$, which we denote by $\sigma (S)$. If the difference of the two resources, $\pi (R)$ and $\sigma (S)$, is a small quantity (i.e., $\Delta^{{\cal D}}(\pi (R),\sigma (S))\le \epsilon$ for small $\epsilon$), we consider that the protocol $\pi$ securely constructs $S$ from $R$. More formally, we define the security as follows. 
\begin{teigi}[\cite{MR11, MT10}]{\rm \label{sim_def}  
For resources $R,S$, we say that a protocol {\it $\pi$ constructs $S$ from $R$ with error $\epsilon\in [0,1]$}, denoted by $R \stackrel{\pi, \epsilon}{\Longrightarrow} S$, if the following two conditions are satisfied:  
\begin{enumerate}
\item  Availability: For the set of all distinguishers ${\cal D}$,  we have 
$\displaystyle \Delta ^{{\cal D}}(\pi(\bot^{E}(R)), \bot^{E}(S))\le \epsilon$, 
where $\bot^{E}$ is the converter which blocks the $E$-interface for distinguishers when it is attached to $R$. 
\item  Security: There exists a simulator $\sigma$ such that, for the set of all distinguishers ${\cal D}$, we have 
$\displaystyle \Delta ^{{\cal D}}(\pi(R), \sigma(S))\le \epsilon$. 
\end{enumerate}  
}
\end{teigi} 

In the above definition, we do not require the condition that the simulator is efficient (i.e., polynomial-time). In other words, the simulator may be inefficient. 

The advantage of the above security definition lies in that a protocol having this kind of security remains to be secure even if it is composed with other protocols. Formally, this can be stated as follows. 
\begin{meidai}[\cite{MR11, MT10}] \label{comp_th}
Let $R,S,T$ and $U$ be resources, and let $\pi,\phi$ be converters (or protocols) such that $R \stackrel{\pi, \epsilon}{\Longrightarrow} S$ and $S \stackrel{\phi, \delta}{\Longrightarrow} T$. Then, we have the following:   
\begin{enumerate}
\item[(1)] $\phi \circ \pi$ satisfies $R \stackrel{\phi \circ \pi, \epsilon+\delta}{\Longrightarrow} T$; 
\item[(2)] $\pi \parallel id$ satisfies $R \parallel U \stackrel{\pi \parallel id, \epsilon}{\Longrightarrow} S\parallel U$; and 
\item[(3)] $id \parallel \pi$ satisfies $U \parallel R \stackrel{id \parallel \pi, \epsilon}{\Longrightarrow} U \parallel S$, 
\end{enumerate}
where $id$ is the trivial converter which makes the interfaces of the subsystem accessible through the interfaces of the combined system. 
\end{meidai}

We note that the first property in Proposition \ref{comp_th} means the security for sequential composition. In addition, as stated in \cite{MR11} three properties in Proposition \ref{comp_th} imply the security for parallel composition in the  following sense: For resources $R,R',S,S'$ and converters $\pi,\phi$ such that $R \stackrel{\pi, \epsilon}{\Longrightarrow} S$ and $R' \stackrel{\phi, \delta}{\Longrightarrow} S'$, $\pi \parallel \phi$ satisfies 
$R \parallel R' \stackrel{\pi \parallel \phi, \epsilon+\delta}{\Longrightarrow} S \parallel S'$.
      
\subsection{Ideal Functionality/Channels} 
In this section, we give several definitions of ideal functionality of resources such as the authenticated channel and key sharing resources which are necessary to discuss in this paper.  
\begin{itemize}
\item Authenticated Channel: An {\it authenticated channel} usable once, denoted by $\LAC\ $, transmits a message (or a plaintext) $m$ from Alice's interface (i.e., $A$-interface) to Bob's interface (i.e., $B$-interface) without any error/replacement. If Eve is active, through the $E$-interface Eve obtains $m$, and she obtains nothing, otherwise. Similarly, an authenticated channel from $B$-interface to $A$-interface can be defined and denoted by $\RAC\ $. For a positive integer $t$, we write $(\LAC\ )^t$ for the composition of invoked $t$ authenticated channels $\LAC\ \| \LAC\ \| \cdots \| \LAC\ $ ($t$ times), and we write $(\LAC\ )^{\infty }$ if arbitrarily many use of $\LAC\ $ is allowed. Similarly, $(\RAC\ )^t$ and $(\RAC\ )^{\infty }$ can be defined. 
\item Secure Channel: A {\it secure channel} usable once, denoted by $\SEC \ $, transmits a plaintext $m$ from $A$-interface to $B$-interface without any error/replacement. Even if Eve is active, she obtains nothing except for the length of the plaintexts and cannot replace $m$ with another plaintext.  Also, for a positive integer $t$, we write $(\SEC\ )^t$ for the composition of invoked $t$ secure channels $\SEC\ \| \SEC\ \| \cdots \| \SEC\ $ ($t$ times). 
\item {Key Sharing Resource (with Uniform Distribution)}: A {\it key sharing resource} with the uniform distribution usable once, denoted by $\KA $, means the ideal resource with no input which generates a uniform random string $k$ and outputs it at both interfaces of Alice and Bob. Even if Eve is active, her interface outputs no information on $k$ and cannot replace $k$ with another one. 
More generally, if such a key $k$ is chosen according to a distribution $P_K$ (not necessarily the uniform distribution), we denote the key sharing resource by [$P_K$].
\item {Correlated Randomness Resource (or Key Distribution Resource)}: Let $P_{XY}$ be a probability distribution with random variables $X$ and $Y$. A {\it correlated randomness resource} usable once, denoted by [$P_{XY}$], means the resource with no input which randomly generates $(x,y)$ according to the distribution $P_{XY}$ and outputs $x$ and $y$ at interfaces of Alice and Bob, respectively. Even if Eve is active, her interface outputs no information on $(x,y)$ and cannot replace it with another one. Note that the resource  [$P_{XY}$] includes [$P_K$] (and hence $\KA\ $) as a special case.  
\end{itemize}
\section{Symmetric-key Encryption: Protocol Execution and Security Formalizations, Revisited}\label{exe_enc}
We explain  the traditional protocol execution of symmetric-key encryption.  
In the following, let ${\cal M}$ and ${\cal C}$ be finite sets of plaintexts and ciphertexts, respectively. In addition, let $M$ and $C$ be andom variables which take plaintexts in ${\cal M}$ and ciphertexts in ${\cal C}$, respectively.    

Let $\pi=(\pi^{A}_{enc},\pi^{B}_{dec})$ be a symmetric-key encryption protocol connected to a key sharing resource [$P_K$] as defined below, where $\pi^{A}_{enc}$ is a converter called an {\it encryption algorithm} at Alice's side, and $\pi^{B}_{dec}$ is a converter called a {\it decryption algorithm} at Bob's side: 
\begin{enumerate}
\item[--] Input of Alice's outer interface: $m \in {\cal M}$
\item[--] {Input of Alice's inner interface}: $k\in {\cal K}$ by accessing [$P_K$]
\item[--] {Input of Bob's inner interface}: $k\in {\cal K}$ by accessing [$P_K$]
\item[--]  {Output of Bob's outer interface}: $\tilde{m}\in \tilde{{\cal M}}$
\item[1.] $\pi^A_{enc}$ computes $c=\pi^A_{enc}(k,m)$ and sends $c$ to $\pi^B_{enc}$ by $\LAC\ $.
\item[2.] $\pi^B_{dec}$ computes $\tilde{m}=\pi^B_{dec}(k,c)$ and outputs $\tilde{m}$.
\end{enumerate}

\medskip

In this paper, for given random variables $M$ and $K$, let $\tilde{M}:=\pi^B_{dec}(K,\pi^A_{enc}(K,M))$ be a random variable which takes values in the set of output of $\pi^{B}_{dec}$.  

Note that we do not require any restriction on the protocol execution of symmetric-key encryption such as: $\pi^A_{enc}$ is deterministic; or for each $k \in {\cal K}$, $\pi^A_{enc}(k,\cdot): {\cal M} \to {\cal C}$ is injective; or $\pi^B_{dec}$ is deterministic; or it has to be satisfied that $\pi^B_{dec}(k,\pi^A_{enc}(k,m))=m$ for any possible $k$ and $m$.    
Therefore, we deal with a general case of the protocol execution of symmetric-key encryption. In particular, it should be noted that: $\pi^{A}_{enc}$ can be probabilistic (i.e., not necessarily deterministic); for each $k\in {\cal K}$, $\pi^{A}_{enc}(k,\cdot)$ may not be injective; $\pi^B_{dec}$ can be probabilistic; and a decryption-error may occur. 

If a symmetric-key encryption protocol $\pi$ is usable at most one time (i.e., the one-time model), the purpose of $\pi$ is to transform the resources $[P_K]$ and $\LAC\ $ into the secure channel $\SEC \ $. In this paper, we only deal with symmetric-key encryption protocols in the one-time model, since this model is simple and fundamental.  

Now, we revisit the formalization of several information-theoretic security notions for symmetric-key encryption. 
The traditional security is formalized based on the notion that the observed ciphertext $C$ and underlying plaintext $M$ are statistically independent. The most famous formalization based on this notion is Shannon's perfect secrecy \cite{Shannon}, $H(M|C)=H(M)$, or equivalently, $I(M;C)=0$. As an extended (or a relaxed) version, we can also consider its variant, $I(M;C)\le \epsilon$ for some small quantity $\epsilon$. Along with this concept, we first consider the following two definitions.    

\medskip

\begin{teigi} {\rm \label{PS}
Let $\pi$ be a symmetric-key encryption protocol. Let $P_M$ be a certain probability distribution on ${\cal M}$. Then, $\pi$ is said to be {\it $\epsilon$-secure for $P_M$} if it satisfies the following conditions:   
\begin{eqnarray*} 
&&\mbox{(i) \ Correctness} \quad \Pr \{ M \neq \tilde{M}\} \le \epsilon; \mbox{ and }\\
&&\mbox{(ii) \ Secrecy} \quad I(M;C)\le \epsilon.  
\end{eqnarray*}
In particular, $\pi$ is said to be {\it perfectly-secure for $P_M$} if $\epsilon=0$ above for $P_M$. 
} 
\end{teigi}

\medskip

\begin{teigi} {\rm \label{UPS}
Let $\pi$ be a symmetric-key encryption protocol. Then, $\pi$ is said to be {\it $\epsilon$-secure}, if for any probability distribution $P_M\in {\cal P}({\cal M})$, we have:  
\begin{eqnarray*}
&&\mbox{(i) \ {Correctness} } \quad \Pr \{ M \neq \tilde{M}\} \le \epsilon; \mbox{ and }\\
&&\mbox{(ii) \ {Secrecy} } \quad I(M;C)\le \epsilon.  
\end{eqnarray*}
In particular, $\pi$ is said to be {\it perfectly-secure} if $\epsilon=0$ above.  
}
\end{teigi}

\medskip

The difference of Definitions \ref{PS} and \ref{UPS} is that we consider security only for a certain distribution of plaintexts or for all distributions of plaintexts. Obviously, Definition \ref{UPS} is stronger than   Definition \ref{PS}, since we can find a distribution $P_M$ and $\pi$ such that $\pi$ is $\epsilon$-secure for $P_M$ but it is not $\epsilon$-secure.  
In this paper, we are interested in correctness and security of Definition \ref{UPS} or other formalizations in which all distributions of plaintexts are considered. From this viewpoint, we give the following definition in a comprehensive way.

\medskip 

\begin{teigi}[Correctness and Security] \label{enc_def_security}
Let $\pi$ be a symmetric-key encryption protocol. Then, $\pi$ is said to be {\it $(\delta, \epsilon)$-secure in the sense of Type $(i,j)$}, if $\pi$ satisfies 
\begin{eqnarray*}
\delta_{\pi,i}\le \delta \mbox{ and } \epsilon_{\pi,j}\le \epsilon, 
\end{eqnarray*}
where $\delta_{\pi,i}$ $(1\le i\le 3)$ and $\epsilon_{\pi,j}$ $(1\le j\le 10)$ are defined as follows. 
\begin{itemize}
\item Correctness. We define the following parameters concerning correctness of $\pi$: 
\begin{eqnarray*}
\delta_{\pi,1}&:=&\sup_{P_M} \Pr \{ M \neq \tilde{M}\}, \\ 
\delta_{\pi,2}&:=&\sup_{P_M} \Delta (P_{M\tilde{M}},P_{MM}), \\ 
\delta_{\pi,3}&:=&\max_{m} \Delta (P_{\tilde{M} | M=m},P_{M|M=m}),   
\end{eqnarray*}
where the supremum ranges over all $P_M\in {\cal P}({\cal M})$.
\item Traditional Secrecy (TS). We define the following advantage of adversaries in terms of traditional secrecy: 
\begin{eqnarray*}
\epsilon_{\pi,1}&:=&\sup_{P_M} I(M;C),\\ 
\epsilon_{\pi,2}&:=&\sup_{P_M} \Delta (P_{MC},P_{M}P_{C}),\\
\epsilon_{\pi,3}&:=&\sup_{P_M} \max_{m\in {\cal M}}\Delta (P_{C|M=m},P_{C}),\\ 
\epsilon_{\pi,4}&:=&\sup_{P_M} \max_{c\in {\cal C}}\Delta (P_{M|C=c},P_{M}).
\end{eqnarray*}
\item Indistinguishability (IND). We define the following advantage of adversaries in terms of indistinguishability:  
\begin{eqnarray*}
\epsilon_{\pi,5}&:=&\max_{m} \max_{m'\not= m} \Delta (P_{C|M=m},P_{C|M=m'}),\\ 
\epsilon_{\pi,6}&:=&\max_{m} \max_{m'\not= m} \ \max_{f:{\cal C}\to \{0,1 \}} \left| \Pr \{f(C)=1 \mid M=m \}- \Pr \{f(C)=1 \mid M=m' \} \right|,
\end{eqnarray*}
where in $\epsilon_{\pi,6}$ the maximum ranges over all functions $f:{\cal C}\to \{0,1 \}$. 
\item Semantic Security (SS). We define the following advantage of adversaries in terms of semantic security: 
\begin{eqnarray}
\epsilon_{\pi,7}:= \sup_{P_M} \ \max_{f:{\cal C}\to \{0,1 \}} \ \inf_{G_f} \ \max_{h:{\cal M}\to \{0,1 \}} \  
\left| \Pr \{ f(C)=h(M) \}- \Pr \{ G_f=h(M) \} \right|, \label{eq:RW_def}
\end{eqnarray}
where the maximum ranges over all functions $f:{\cal C}\to \{0,1 \}$ and $h:{\cal M}\to \{0,1 \}$, and the infimum ranges over all binary random variable $G_f$ which only depends on $f$ but is independent of $P_M$ and $h$.
\item Composable Security (CS). We define the following advantage of adversaries in terms of composable security: 
\begin{eqnarray}
\epsilon_{\pi,8}&:=&\inf_{\sigma} \Delta^{{\cal D}}(\pi (\LAC \ ||\ [\mbox{$P_{K}$}]),\sigma (\SEC \ )) \label{UC8_1}\\
&=& \inf_{P_{Q}} \sup_{P_M} \Delta (P_{M \tilde{M} C},P_{M M}P_{Q}), \label{UC8_2} \\
\epsilon_{\pi,9}&:=&\inf_{P_Q} \sup_{P_M} \Delta (P_{MC},P_{M}P_Q), \label{UC9}\\
\epsilon_{\pi,10}&:=&\inf_{P_Q} \max_{m} \Delta (P_{C|M=m},P_Q),  \label{UC10}
\end{eqnarray}
where the infimum in (\ref{UC8_1}) ranges over all possible simulators, the supremum in (\ref{UC8_2}), (\ref{UC9}) ranges over all $P_M\in {\cal P}({\cal M})$, and the infimum in (\ref{UC8_2})--(\ref{UC10}) ranges over all $P_Q\in {\cal P}({\cal C})$.
\end{itemize} 
\end{teigi}

By Definition \ref{enc_def_security}, we can formally give thirty kinds of formalizations of correctness and security. 
In particular, several important formalizations known can be considered to be captured within Definition \ref{enc_def_security} as follows. 
\begin{itemize}
\item Traditional Secrecy (TS). The traditional formalization in Definition \ref{UPS} corresponds to the security in the sense of Type $(1,1)$. Note that, instead of using the mutual information, independence of $M$ and $C$ is expressed by the statistical distance as $\Delta (P_{MC},P_MP_C)=0$, and can be relaxed as $\Delta (P_{MC},P_MP_C)\le \epsilon$. This type of security is represented by Type $(1,2)$. 
\item Indistinguishability (IND). The formalization based on information-theoretic analogue of indistinguishability by Goldwasser and Micali \cite{GM} corresponds to the security in the sense of Type $(1,5)$, since $\epsilon_{\pi,5}$ means the adversary's advantage for distinguishing the views (i.e., distributions of ciphertexts) in the protocol execution when two different plaintexts are inputted. In addition, $\epsilon_{\pi,6}$ means another interpretation of information-theoretic indistinguishability, since the adversary's advantage for distinguishing the views is described by the use of a binary function $f$ arbitrarily chosen by the adversary. In this case, the security is represented by Type $(1,6)$.   
\item Semantic Security (SS).  The formalization based on information-theoretic analogue of semantic security by Goldwasser and Micali \cite{GM} corresponds to the security in the sense of Type $(1,7)$ by the following reason: 
Intuitively, semantic security implies that a ciphertext $C$ is almost useless to obtain any one bit information of the underlying plaintext $M$; and the adversary's advantage $\epsilon_{\pi,7}$ implies that, in order to guess such one bit information $h(M)$, there is no difference between by using the ciphertext $C$ and a mapping $f$, and by using $f$ only with a random coin. 
\item Composable Security (CS). The formalizations based on information-theoretic composable security given by Definition \ref{sim_def} is the security in the sense of Type $(2,8)$ or Type $(3,8)$. In addition, we can consider the distinguisher's advantage by the following his behavier: a distinguisher arbitrarily chooses a random variable $M$ (or a plaintext $m$) and inputs it into $A$-interface; then, $\delta_{\pi,2}$ (or $\delta_{\pi,3}$) means the distinguisher's advantage for distinguishing real output and ideal one at $B$-interface; and $\epsilon_{\pi,9}$ (or $\epsilon_{\pi,10}$) means his  advantage for distinguishing real output and simulator's output (according to $P_Q$) at $E$-interface. By combining those, we will reach the security of Type $(i,j)$ with $i=2,3$ and $j=9,10$ for the composable security.    
\end{itemize}   

We next define equivalence of security notions of Type $(i,j)$ as follows. 

\medskip

\begin{teigi}  
For a symmetric-key encryption protocol $\pi$, its security of Type $(i,j)$ and Type $(i',j')$ are said to be {\it strictly equivalent}, if $\delta_{\pi,i}=\Theta (\delta_{\pi,i'})$ and $\epsilon_{\pi,j}=\Theta (\epsilon_{\pi,j'})$ where $\Theta (\cdot)$ is evaluated in a system parameter $\kappa \in \mathbb{N}$ inputted into $\pi$ and [$P_K$].    
Furthermore, for a symmetric-key encryption protocol $\pi$, its security of Type $(i,j)$ and Type $(i',j')$ are said to be {\it equivalent}, if it holds that  
\begin{eqnarray*}
 (\delta_{\pi,i}, \epsilon_{\pi,j})\to (0,0) \mbox{ (as $\kappa\to \infty $}) \quad \mbox{if and only if} \quad (\delta_{\pi,i'}, \epsilon_{\pi,j'}) \to (0,0)  \mbox{ (as $\kappa\to \infty $}).
\end{eqnarray*}
\end{teigi}

In Section \ref{relation_enc}, we will show equivalence and non-equivalence between the formalizations in a comprehensive way.

\section{Symmetric-key Encryption: Relationships between Formalizations of Correctness and Security}\label{relation_enc}
\subsection{Equivalence}\label{Eq_relation}
We show the explicit relationships between security formalizations of Type $(i,j)$ for $1\le i \le 3$ and $1\le j \le 10$. In the following, let $\Pi$ be a family of all symmetric-key encryption protocols. 

\medskip

\begin{teiri}\label{main1} 
For any symmetric-key encryption protocol $\pi \in \Pi$, we have explicit relationships between formalizations of correctness and security as follows.
\begin{enumerate}
\item Correctness formalizations: $\delta_{\pi,1}=\delta_{\pi,2}=\delta_{\pi,3}$, 
\item TS formalizations: 
\begin{eqnarray*}
\frac{2}{\ln 2}\epsilon_{\pi,2}^2 \le \epsilon_{\pi,1}\le - 2\epsilon_{\pi,2} \log \frac{2\epsilon_{\pi,2}}{|{\cal M}|\ |{\cal C}|}, \quad \epsilon_{2,\pi}\le \epsilon_{3,\pi}\le 2 \epsilon_{2,\pi}, \quad \epsilon_{2,\pi}\le \epsilon_{4,\pi},  
\end{eqnarray*}
\item IND formalizations: $\epsilon_{5,\pi}= \epsilon_{6,\pi}$, 
\item CS formalizations: $\max (\epsilon_{\pi,9},\delta_{\pi,2}) \le \epsilon_{\pi,8} \le \epsilon_{\pi,9}+\delta_{\pi,2}, \quad \epsilon_{9,\pi}= \epsilon_{10,\pi}$,
\item TS and IND: $\epsilon_{3,\pi}= \epsilon_{5,\pi} $, 
\item SS and IND: $\epsilon_{\pi,7} \le \epsilon_{\pi,6} \le 4 \epsilon_{\pi,7}$, 
\item IND, CS, and TS: $\frac{1}{2}\epsilon_{2,\pi} \le \epsilon_{9,\pi}\le \epsilon_{5,\pi}$.
\end{enumerate} 
\end{teiri}

\medskip

{\it Proof of Theorem \ref{main1}.}
The proof is organized as follows:  
\begin{enumerate}
\item Proof of relationships between correctness formalizations is given by Lemma \ref{Relation_correctness},
\item Proof of relationships between TS formalizations is given by Lemmas \ref{Relation_TS} and \ref{Relation_TS_last},
\item Proof of relationships between IND formalizations is given by Lemma \ref{Relation_IND},  
\item Proof of relationships between CS formalizations is given by Lemmas \ref{Relation_CS1} and \ref{Relation_CS2},
\item Proof of relationships between TS and IND is given by Lemma \ref{TS/IND}, 
\item Proof of relationships between SS and IND is given by Lemma \ref{Relation_IND/SS},
\item Proof of relationships among IND, CS, and TS is given by Lemma \ref{TS_CS_IND}.  
\end{enumerate}
In the following, we will show Lemmas 1--9 to complete the proof. 

\medskip

\begin{hodai}\label{Relation_correctness}
For any symmetric-key encryption $\pi$, we have 
$\delta_{\pi,1}=\delta_{\pi,2}=\delta_{\pi,3}$.
\end{hodai}
\begin{IEEEproof} 
First, we show $\delta_{\pi,1}=\delta_{\pi,2}$: For any $\pi$ and for any distribution $P_M$, we have $\Delta (P_{MM},P_{M\tilde{M}})=P(M \neq \tilde{M})$ by Proposition \ref{hosoku2} in Appendix A, from which it is straightforward to have $\delta_{\pi,1}=\delta_{\pi,2}$.   

Secondly, we show $\delta_{\pi,2}=\delta_{\pi,3}$: This is shown by applying $X=Y=M$ and $Z=\tilde{M}$ in Proposition \ref{hosoku_relation1} in Appendix B. 
\end{IEEEproof}

\begin{hodai}\label{Relation_TS} 
For any symmetric-key encryption $\pi$, we have 
\begin{eqnarray*}
&& \frac{2}{\ln 2}\epsilon_{\pi,2}^2 \le \epsilon_{\pi,1}\le - 2\epsilon_{\pi,2} \log \frac{2\epsilon_{\pi,2}}{|{\cal M}|\ |{\cal C}|},\\ 
&& \epsilon_{\pi,2}\le \epsilon_{\pi,3}, \mbox{ and } \epsilon_{\pi,2} \le \epsilon_{\pi,4}.
\end{eqnarray*}
\end{hodai} 
\begin{IEEEproof} 
First, we show that $\frac{2}{\ln 2}\epsilon_{\pi,2}^2 \le \epsilon_{\pi,1}\le - 2\epsilon_{\pi,2} \log \frac{2\epsilon_{\pi,2}}{|{\cal M}|\ |{\cal C}|}$:  
From Corollary \ref{CT3} in Appendix A, it follows that, for any $P_M$ and any $\pi$,   
\begin{eqnarray*}
I(M;C)&\le& - 2\Delta (P_{MC},P_{M}P_{C}) \log \frac{2\Delta (P_{MC},P_{M}P_{C})}{|{\cal M}|\cdot|{\cal C}|}\\
&\le& - 2\epsilon_{\pi,2} \log \frac{2\epsilon_{\pi,2}}{|{\cal M}|\cdot|{\cal C}|}.  
\end{eqnarray*}
Therefore, we have 
\begin{eqnarray*}
\epsilon_{\pi,1}\le - 2\epsilon_{\pi,2} \log \frac{2\epsilon_{\pi,2}}{|{\cal M}|\cdot|{\cal C}|}. 
\end{eqnarray*}

On the other hand, from Corollary \ref{CT4} in Appendix A, it follows that, for any $P_M$ and any $\pi$, 
\begin{eqnarray*}
\Delta (P_{MC},P_{M}P_{C})\le \sqrt{\frac{\ln 2}{2}} I(M;C)^{\frac{1}{2}}. 
\end{eqnarray*}
Therefore, we have $\epsilon_{\pi,2}\le \sqrt{\frac{\ln 2}{2}}\epsilon_{\pi,1}^{\frac{1}{2}}$. 

Secondly, we show $\epsilon_{\pi,2}\le \epsilon_{\pi,3}$: 
For an arbitrary distribution $P_M$, we have 
\begin{eqnarray*}
\Delta (P_{M C},P_{M}P_C)
&=&\frac{1}{2}\sum_{m,c}|P_{M C}(m,c)-P_{M}(m)P_{C}(c) |\\
&= &\frac{1}{2} \sum_{m} P_{M}(m) \sum_{c} |P_{C|M=m}(c|m)- P_{C}(c)|\\
&\le&\frac{1}{2} \max_{m} \sum_{c} |P_{C|M=m}(c|m)- P_{C}(c)|  \\
&= & \max_{m} \Delta (P_{C|M=m}, P_{C}).
\end{eqnarray*}
Therefore, we get $\epsilon_{\pi,2} \le \epsilon_{\pi,3}$. 

Similarly, we can show that, for an arbitrary distribution $P_M$, 
\begin{eqnarray*}
\Delta (P_{M C},P_{M}P_C)\le \max_{c} \Delta (P_{M|C=c}, P_{M}),
\end{eqnarray*}
which implies that $\epsilon_{\pi,2} \le \epsilon_{\pi,4}$.  
\end{IEEEproof}

\medskip
 
\begin{hodai}\label{Relation_IND} 
For any symmetric-key encryption $\pi$, we have 
$\epsilon_{\pi,5}=\epsilon_{\pi,6}$.
\end{hodai}
\begin{IEEEproof} 
For probability distributions $P_X$ and $P_Y$ over a finite set ${\cal X}$, it holds that 
\begin{eqnarray*}
\Delta (P_X,P_Y)=\max_{f:{\cal X}\to \{0,1\}} \left| \Pr \{ f(X)=1 \} - \Pr \{ f(Y)=1 \} \right|.
\end{eqnarray*}
Thus, we have 
\begin{eqnarray*}
\max_{m,m'}\Delta (P_{C|M=m},P_{C|M=m'})
=\max_{m,m'}\max_{f:{\cal C}\to \{0,1\}} \left| \Pr \{ f(C)=1 \mid M=m \} - \Pr \{ f(C)=1 \mid M=m' \} \right|,
\end{eqnarray*}
which implies $\epsilon_{\pi,5}=\epsilon_{\pi,6}$.
\end{IEEEproof} 

\medskip

\begin{hodai}\label{Relation_CS1}
For any symmetric-key encryption $\pi$, we have 
\begin{eqnarray*}
\max (\epsilon_{\pi,9},\delta_{\pi,2}) \le \epsilon_{\pi,8} \le \epsilon_{\pi,9}+\delta_{\pi,2}. 
\end{eqnarray*}
\end{hodai}
\begin{IEEEproof}
For any distributions $P_M\in {\cal P}({\cal M})$ and $P_Q\in {\cal P}({\cal C})$, we have 
\begin{eqnarray*}
\Delta (P_{M \tilde{M} C},P_{M M}P_{Q}) 
&\le& \Delta (P_{M \tilde{M} C},P_{M M C})+ \Delta (P_{M M C},P_{M M}P_Q) \\
&=&  \Delta (P_{M \tilde{M}},P_{M M}) + \Delta (P_{M C},P_{M}P_Q). 
\end{eqnarray*} 
By taking the supremum over all $P_M\in {\cal P}({\cal M})$ and the infimum over all $P_Q\in {\cal P}({\cal C})$, we have 
\begin{eqnarray*}
\inf_{P_Q} \sup_{P_M} \Delta (P_{M \tilde{M} C},P_{M M}P_{Q}) 
&\le& \sup_{P_M}  \Delta (P_{M \tilde{M}},P_{M M}) + \inf_{P_Q} \sup_{P_M} \Delta (P_{M C},P_{M}P_Q)\\
&=& \delta_{\pi,2}+\epsilon_{\pi,9}. 
\end{eqnarray*} 

In addition, from Proposition \ref{hosoku1} in Appendix A, it is clear that $\Delta (P_{M C},P_{M}P_Q) \le \Delta (P_{M \tilde{M} C},P_{M M}P_{Q})$ for any $P_M\in {\cal P}({\cal M})$ and $P_Q\in {\cal P}({\cal C})$. Therefore, we obtain 
\begin{eqnarray*}
\epsilon_{\pi,9}\le \inf_{P_Q} \sup_{P_M} \Delta (P_{M \tilde{M} C},P_{M M}P_{Q}). 
\end{eqnarray*}
Similarly, we have $\delta_{\pi,2}\le \inf_{P_Q} \sup_{P_M} \Delta (P_{M \tilde{M} C},P_{M M}P_{Q})$. 
\end{IEEEproof}

\medskip

\begin{hodai}\label{Relation_CS2}
For any symmetric-key encryption $\pi$, we have $\epsilon_{\pi,9}=\epsilon_{\pi,10}$. 
\end{hodai}
\begin{IEEEproof} 
For arbitrary distributions $P_Q$ and $P_M$, we set $X:=M$, $Y:=C$, and $Z:=Q$, and use Proposition \ref{hosoku_relation1} in Appendix B. Then, we have 
$\sup_{P_M}\Delta(P_{MC},P_M P_Q)=\max_{m}\Delta(P_{C|M=m},P_Q)$. 
Therefore, by taking the infimum over all $P_Q\in {\cal P}({\cal C})$, we have $\epsilon_{\pi,9}=\epsilon_{\pi,10}$. 
\end{IEEEproof}

\medskip

\begin{hodai}\label{TS/IND}
For any symmetric-key encryption $\pi$, we have $\epsilon_{\pi,3}= \epsilon_{\pi,5}$. 
\end{hodai} 
\begin{IEEEproof}
Observe for every $m\in\c{M}$ that 
\begin{eqnarray}
\Delta \left(P_{C|M=m},P_C \right)
&=&\frac{1}{2}\sum_{c \in \c{C}}\left|
P_{C|M}(c|m) - \sum_{m' \in \c{M}} P_{C|M}(c|m')P_M(m')
\right|
\nonumber \\
&=&\frac{1}{2}\sum_{c \in \c{C}}\left|
\sum_{m'\in \c{M}}P_M(m')\left\{P_{C|M}(c|m) - P_{C|M}(c|m')\right\}
\right|. 
\label{eq:INS/PS}
\end{eqnarray}

First, we prove $\epsilon_{\pi,3}\le \epsilon_{\pi,5}$: For arbitrary $P_M$, let $m_0:=\arg \max_{m} \Delta(P_{C|M=m},P_C)$. Then,  from \eqref{eq:INS/PS} we have 
\begin{eqnarray}
\nonumber
\max_{m} \Delta(P_{C|M=m},P_C)&=&\Delta \left(P_{C|M=m_0},P_C \right)\\
\nonumber
&\le&
\frac{1}{2}\sum_{m_1\in\c{M}}P_M(m_1)\sum_{c \in \c{C}}\left|
P_{C|M}(c|m_0) - P_{C|M}(c|m_1)\right|\\
\nonumber
&=& \sum_{m_1\in \c{M}} P_M(m_1) \cdot \Delta \bigl(P_{C|M=m_0}, P_{C|M=m_1} \bigr)\\
\nonumber
&\le& \sum_{m_1 \in \c{M}} P_M(m_1) \epsilon_{\pi,5}\\
&=& \epsilon_{\pi,5}.
\end{eqnarray}
Hence, by taking the supremum over all $P_M\in {\cal P}({\cal M})$, we get $\epsilon_{\pi,3}\le \epsilon_{\pi,5}$.

Next, we show $\epsilon_{\pi,3}\ge \epsilon_{\pi,5}$: Let $m_0,m_1\in {\cal M}$ such that $\epsilon_{\pi,5}=\Delta(P_{C|M=m_0},P_{C|M=m_1})$. For arbitrary $\epsilon>0$, we define 
\begin{eqnarray}\label{eq:delta_f}
P_M(m')=\left\{
\begin{array}{cll}
1-\gamma,&\mbox{if}~~m'=m_1\\
\frac{\gamma}{|{\cal M}|-1},&\mbox{otherwise}
\end{array}
\right. 
\end{eqnarray}
where $\gamma$ is a positive real number such that $\gamma \epsilon_{\pi,5} \le \epsilon$. 
Then, by substituting both $m=m_0$ and $P_M(m')$ into \eqref{eq:INS/PS}, we obtain
\begin{eqnarray*}
\epsilon_{\pi,3}
&\ge&\Delta(P_{C|M=m_0},P_C) \\
&=& \frac{1}{2}\sum_{c \in \c{C}}\left|
\sum_{m'\in \c{M}}P_M(m')\left\{P_{C|M}(c|m_0) - P_{C|M}(c|m')\right\} \right| \\
&\ge& \frac{1}{2}\sum_{c \in \c{C}}\left|
(1-\gamma) \left\{P_{C|M}(c|m_0) - P_{C|M}(c|m_1)\right\} \right| \\
&=&(1-\gamma)\epsilon_{\pi,5}\\
&\ge& \epsilon_{\pi,5}-\epsilon. 
\end{eqnarray*}  
\end{IEEEproof}

\medskip

\begin{hodai}\label{Relation_IND/SS}
For any symmetric key encryption $\pi$, we have 
\begin{eqnarray*}
\epsilon_{\pi,7} \le \epsilon_{\pi,6} \le 4 \epsilon_{\pi,7}.  
\end{eqnarray*}
\end{hodai}
\begin{IEEEproof}
First, we prove $\epsilon_{\pi,7} \le \epsilon_{\pi,6}$. 
This part of the proof can be shown in a very similar way as that in \cite{G_FCv1-2001} as follows, though the proof in \cite{G_FCv1-2001} is given under computational security setting. 
Suppose that a distribution $P_M$ and functions $f:{\cal C}\to \{0,1\}$, $h:{\cal M}\to \{0,1\}$ are arbitrarily given.  
Let $P_{M^*}$ be an independently and identically distribution of $P_M$. Then, we consider the random variable $G_f$ which is defined by 
\begin{eqnarray*}
G_f := f(C^*), \mbox{ and } P_{C^*}(c) := \sum_{m_1}P_{C|M}(c|m_1)P_{M^*}(m_1) \mbox{ for $c\in\c{C}$ and $m_1 \in \c{M}$}.
\end{eqnarray*}

Let us define an indicator function $\textbf{1}_{f,h}:\c{C} \times \c{M} \rightarrow \{0,1\}$ for maps $f$ and $h$ by  
\begin{eqnarray}
\textbf{1}_{f,h}(c,m) = 
\left\{
\begin{array}{cll}
1, & \mbox{if}~~f(c)=h(m)\\
0, & \mbox{otherwise}.\\
\end{array}
\right.
\end{eqnarray}
Then, we have 
\begin{eqnarray*}
&&\inf_{G_{\hat{f}}}\left| \Pr \{f(C)=h(M)\} - \Pr \{G_{\hat{f}}=h(M) \} \right| \\
&&\le \left| \Pr \{f(C)=h(M)\} - \Pr \{G_f=h(M) \} \right| \\
&&= \left| \Pr \{f(C)=h(M)\} - \Pr \{ f(C^*)=h(M) \} \right| \\
&&= \left| \sum_{c,m_0} \textbf{1}_{f,h}(c,m_0) \left\{ P_{CM}(c,m_0) - P_{C^*M}(c,m_0) \right\} \right| \\
&&= \left| \sum_{c,m_0} \textbf{1}_{f,h}(c,m_0)P_{M}(m_0) \left\{ P_{C|M}(c|m_0) - P_{C^*}(c) \right\} \right| \\
&&= \left| \sum_{m_0,m_1} P_{M}(m_0)P_{M^*}(m_1) \sum_{c} \textbf{1}_{f,h}(c,m_0)\{P_{C|M}(c|m_0) - P_{C|M}(c|m_1) \} \right| \\
&&= \left| \sum_{m_0,m_1} P_{M}(m_0)P_{M^*}(m_1) 
\left\{ \Pr \{f_{h,m_0}(C)=1|M=m_0 \} -\Pr \{f_{h,m_0}(C)=1|M=m_1 \} \right\} \right| \label{eq:IND->SS}\\
&&\le  \sum_{m_0,m_1} P_{M}(m_0)P_{M^*}(m_1) \epsilon_{\pi,6}\\
&&=\epsilon_{\pi,6},
\end{eqnarray*}
where $f_{h,m_0}: \c{C} \rightarrow \{0,1\}$ is defined by $f_{h,m_0}(c)=1$ if and only if $\textbf{1}_{f,h}(c,m_0)=1$. 
Therefore, we have $\epsilon_{\pi,7} \le \epsilon_{\pi,6}$. 

Next, we show that $\epsilon_{\pi,6} \le 4 \epsilon_{\pi,7}$. 
We first prove the following claim. 

\medskip

\begin{claim}
For arbitrarily given $P_M$, $f:\c{C} \rightarrow \{0,1\}$, and $h:\c{M} \rightarrow \{0,1\}$, we have 
\begin{eqnarray}
\biggr| \Pr \{ f(C)=h(M) \}-\sum_{\ell \in \{0,1\}} \Pr \{ f(C)=\ell \} \Pr \{ h(M)=\ell \}\biggr|\le 2 \epsilon_{\pi,7}.\label{relation_SS_0}
\end{eqnarray}
\end{claim}
\begin{IEEEproof}
Suppose that $P_M$ and $f:\c{C} \rightarrow \{0,1\}$ are arbitrarily given. Then, by definition of semantic security, there exists $G_f$ such that 
\begin{eqnarray}
\biggr| \Pr \{ f(C)=h(M) \}- \Pr \{ G_f=h(M) \} \biggr| \le \epsilon_{\pi,7}\label{relation_SS}
\end{eqnarray} 
for any $h:\c{M} \rightarrow \{0,1\}$. In particular, letting $h$ be a map that always outputs $1$ for every $m\in\c{M}$, we have  
\begin{eqnarray*}
\bigl| \Pr \{ f(C) = 1 \} - \Pr \{ G_f = 1 \} \bigr| \le \epsilon_{\pi,7}
\end{eqnarray*}
which is equivalent to
\begin{eqnarray*}
\bigl| \Pr \{ f(C) = 0 \} - \Pr \{ G_f = 0 \} \bigr| \le \epsilon_{\pi,7}.
\end{eqnarray*}
Thus, for $\ell\in\{0,1\}$, it holds that 
\begin{eqnarray*}
\Pr \{ G_f=\ell \}+\epsilon_{\pi,7} \ge \Pr \{ f(C)=\ell \} \ge \Pr \{ G_f=\ell \}-\epsilon_{\pi,7},
\end{eqnarray*}
and hence we have 
\begin{eqnarray*}
(\Pr \{ G_f=\ell \}+\epsilon_{\pi,7})\Pr \{ h(M)=\ell \} 
\ge \Pr \{ f(C)=\ell \}\Pr \{ h(M)=\ell \} 
\ge (\Pr \{ G_f=\ell \}-\epsilon_{\pi,7})\Pr \{ h(M)=\ell \}.
\end{eqnarray*}
From this, it follows that
\begin{eqnarray*}
\Pr \{ G_f=h(M) \}+\epsilon_{\pi,7}
\ge \sum_{\ell \in \{ 0,1 \} } \Pr \{ f(C)=\ell \}\Pr \{ h(M)=\ell \} 
\ge \Pr \{ G_f=h(M) \}-\epsilon_{\pi,7}, 
\end{eqnarray*}
or equivalently, 
\begin{eqnarray}
\biggr|  \sum_{\ell \in \{ 0,1 \} } \Pr \{ f(C)=\ell \}\Pr \{ h(M)=\ell \} -  \Pr \{ G_f=h(M) \} \biggr| \le \epsilon_{\pi,7}.\label{relation_SS_1}
\end{eqnarray}
Therefore, we obtain 
\begin{eqnarray*}
&&\biggr| \Pr \{ f(C)=h(M) \}-\sum_{\ell \in \{0,1\}} \Pr \{ f(C)=\ell \} \Pr \{ h(M)=\ell \}\biggr| \\
&&\le \biggr| \Pr \{ f(C)=h(M) \}- \Pr \{ G_f=h(M) \}  \biggr| \\
&& \qquad + \biggr|  \sum_{\ell \in \{ 0,1 \} } \Pr \{ f(C)=\ell \}\Pr \{ h(M)=\ell \} -  \Pr \{ G_f=h(M) \} \biggr|\\
&& \le 2 \epsilon_{\pi,7},
\end{eqnarray*}
where the last inequality follows from (\ref{relation_SS}) and (\ref{relation_SS_1}). 
\end{IEEEproof}

\medskip

By applying $X=f(C)$ and $Y=h(M)$ in Lemma \ref{lem:binary} in Appendix B to the inequality (\ref{relation_SS_0}), it holds that
\begin{eqnarray}
\Bigr| \Pr \{ f(C)=h(M)=1 \} -\Pr \{ f(C)=1 \} \Pr \{ h(M)=1 \} \Bigr|\le \epsilon_{\pi,7}, \label{relation_SS_2}
\end{eqnarray}
for arbitrarily given $P_M$, $f:\c{C} \rightarrow \{0,1\}$, and $h:\c{M} \rightarrow \{0,1\}$. 
In particular, we choose $m_0,m_1\in {\cal M}$ by which $\epsilon_{\pi,6}$ is given. Then, for arbitrary $\epsilon >0$, we consider   
\begin{eqnarray*}
P_{M}(m):=\left\{
  \begin{array}{l}
    \frac{1}{2}, \quad\quad \ \ \mbox{if $m=m_0$},    \\
    \frac{1}{2}-\gamma, \quad \mbox{if $m=m_1$}, \\
    \frac{\gamma}{|{\cal M}|-2}, \quad \mbox{otherwise},  \\
  \end{array}
\right.
\end{eqnarray*}
where $\gamma$ is a positive real number with $\gamma \epsilon_{\pi,6}\le 2\epsilon$,  
and take $h:\c{M} \rightarrow \{0,1\}$ defined by 
\begin{eqnarray*}
h(m):=\left\{
  \begin{array}{l}
    1, \quad \mbox{if $m=m_0$},    \\
    0, \quad \mbox{otherwise}.  \\
  \end{array}
\right.
\end{eqnarray*}
Then, from (\ref{relation_SS_2}) it follows that   
\begin{eqnarray*}
\epsilon_{\pi,7}&\ge& \Bigr| \Pr \{ f(C)=h(M)=1 \} -\Pr \{ f(C)=1 \} \Pr \{ h(M)=1 \} \Bigr|\\
&=&\Pr \{ M=m_0 \} \biggr| \Pr \{ f(C)=1 \mid M=m_0 \} -\sum_{\ell\in\{0,1\}} \Pr \{ f(C)=1\mid M=m_\ell \} \Pr \{ M=m_\ell \}\biggr| \\
&=&\Pr \{ M=m_0 \} \Pr \{ M=m_1 \} \biggr| \Pr \{ f(C)=1 \mid M=m_0 \} - \Pr \{ f(C)=1\mid M=m_1\} \biggr| \\
&=&\frac{1}{2}\left( \frac{1}{2}-\gamma \right) \epsilon_{\pi,6} \\
&\ge&\frac{1}{4}\epsilon_{\pi,6}-\epsilon. 
\end{eqnarray*}
\end{IEEEproof}

\medskip

\begin{hodai}\label{TS_CS_IND}
For any symmetric-key encryption $\pi$, we have $\frac{1}{2}\epsilon_{\pi,2}\le \epsilon_{\pi,9}\le \epsilon_{\pi,5}$. 
\end{hodai}
\begin{IEEEproof}
First, we show $\frac{1}{2}\epsilon_{\pi,2}\le \epsilon_{\pi,9}$: For arbitrary distributions $P_Q$ and $P_M$, we have 
\begin{eqnarray*}
\Delta (P_{MC},P_{M}P_{C}) &\le& \Delta (P_{MC},P_{M}P_Q)+\Delta (P_{M}P_Q,P_{M}P_{C})\\
&=&\Delta (P_{MC},P_{M}P_Q)+\Delta (P_Q,P_{C})\\
&\le&2\Delta (P_{MC},P_{M}P_Q).
\end{eqnarray*}
Therefore, $\epsilon_{\pi,2}\le 2 \alpha_{\pi,9}$.

Next, we show $\epsilon_{\pi,9}\le \epsilon_{\pi,5}$: By Lemma \ref{Relation_CS2}, it is sufficient to prove $\epsilon_{\pi,10}\le \epsilon_{\pi,5}$.  
Let $m_0\in {\cal M}$ be a plaintext such that it gives $\epsilon_{\pi,10}$, and set $P_Q:=P_{C|M=m_1}$ by choosing $m_1\in {\cal M}$ with $m_1\not=m_0$. Then, we have
\begin{eqnarray*}
\epsilon_{\pi,10}\le \Delta (P_{C|M=m_0},P_Q)
= \Delta (P_{C|M=m_0},P_{C|M=m_1})
\le \epsilon_{\pi,5}. 
\end{eqnarray*}
\end{IEEEproof} 

\medskip

\begin{hodai}\label{Relation_TS_last}
For any symmetric-key encryption $\pi$, we have $\epsilon_{\pi,3}\le 2\epsilon_{\pi,2}$. 
\end{hodai}
\begin{IEEEproof}
By Lemma \ref{TS/IND}, it is sufficient to prove  $\epsilon_{\pi,5}\le 2\epsilon_{\pi,2}$. 
For any $\epsilon>0$, and for $m_0,m_1\in {\cal M}$ ($m_0 \not= m_1$) such that $\epsilon_{\pi,5}=\Delta (P_{C|M=m_0},P_{C|M=m_1})$, we define a distribution $P_{\hat{M}}$ by 
\begin{eqnarray*}
P_{\hat{M}}(m):=
\left\{
  \begin{array}{l}
    \frac{1}{2}(1-\gamma) \quad \mbox{  if $m\in \{m_0,m_1 \}$},   \\
    \frac{\gamma}{|{\cal M}|-2} \qquad \mbox{  otherwise},  \\
  \end{array}
\right.
\end{eqnarray*}
where $\gamma$ is a positive real number such that $\gamma \epsilon_{\pi,5} \le 2\epsilon$. Then, we have
\begin{eqnarray*}
\epsilon_{\pi,2}&\ge& \Delta (P_{\hat{M}\hat{C}},P_{\hat{M}}P_{\hat{C}})\\
&\ge&\frac{1}{2}(1-\gamma)\{ \Delta (P_{\hat{C}|\hat{M}=m_0},P_{\hat{C}})+\Delta (P_{\hat{C}|\hat{M}=m_1},P_{\hat{C}}) \}\\
&\ge& \frac{1}{2}(1-\gamma) \Delta (P_{\hat{C}|\hat{M}=m_0},P_{\hat{C}|\hat{M}=m_1})\\
&=&\frac{1}{2}(1-\gamma)\alpha_{\pi,5}\\
&\ge& \frac{1}{2}\epsilon_{\pi,5}-\epsilon. 
\end{eqnarray*}
\end{IEEEproof}

\medskip

The following theorem shows equivalence between security formalizations of Type $(i,j)$ under a certain condition. 

\medskip
  
\begin{teiri}\label{main2}
For security formalizations of Type $(i,j)$ with $1\le i\le 3$ and $1\le j\le 10$, we have the following relationships: 
\begin{enumerate}
\item[(i)] For arbitrary symmetric-key encryption protocol $\pi\in \Pi$, all $\pi$'s security of Type $(i,j)$ are strictly equivalent except for $j=1,4$.
\item[(ii)] Let ${\Pi}_1=\{ \pi\in \Pi \mid \epsilon_{\pi,5}=o(1/\log |{\cal M}|) \mbox{ and } \epsilon_{\pi,5}=o(1/\log |{\cal C}|) \}$. Then, for arbitrary $\pi\in \Pi_1$, all $\pi$'s security of Type $(i,j)$ are equivalent except for $j=4$.
\item[(iii)] Let ${\Pi}_2=\{ \pi\in \Pi \mid |{\cal C}|= \Theta (|{\cal M}|), \ \delta_{\pi,1}=o(1/\log |{\cal M}|), \  \epsilon_{\pi,5}=o(1/|{\cal M}|) \}$. Then, for arbitrary $\pi\in \Pi_2$, all $\pi$'s security of Type $(i,j)$ are equivalent.
\end{enumerate} 
\end{teiri}
\begin{IEEEproof}
First, the proof of (i) directly follows from Theorem \ref{main1}.  

Next, we prove (ii). By Theorem \ref{main1}, we have 
\begin{eqnarray}
\epsilon_{\pi,1}\le - 2\epsilon_{\pi,2} \log \frac{2\epsilon_{\pi,2}}{|{\cal M}|\ |{\cal C}|}, \  \mbox{ and }\epsilon_{\pi,2}=\Theta (\epsilon_{\pi,5}). \label{proof_gap1_siki1}
\end{eqnarray}

Now, we consider the following proposition. 
\begin{hodai} \label{proof_gap1_lem1}
Let $y(x)$ be a continuous and positive function defined over $(0, \infty )$, and  
define the function $f(x)=-x \log \frac{x}{y(x)}$. Then, it holds that $f(x)\to 0$ as $x \to 0$ if and only if $y(x)=2^{o(1/x)}$. 
\end{hodai}
\begin{IEEEproof} 
Note that $f(x)=-x \log x +x\log y(x)$ and $-x \log x \to 0$ as $x\to 0$. Therefore, we have $f(x)\to 0$ if and only if $x\log y(x) \to 0$, or equivalently, $y(x)=2^{o(1/x)}$. 
\end{IEEEproof}

Therefore, by Lemma \ref{proof_gap1_lem1} and (\ref{proof_gap1_siki1}), we have  $\epsilon_{\pi,1}=o(1)$ under the condition $|{\cal M}|\cdot |{\cal C}|=2^{o(1/\epsilon_{\pi,5})}$, which in turn follows from the condition that $\epsilon_{\pi,5}=o(1/\log |{\cal M}|)$ and $\epsilon_{\pi,5}=o(1/\log |{\cal C}|)$.

Finally, in order to prove (iii), we need the following propositions.  
\begin{hodai}\label{Lem_gap1}
Let $\displaystyle \tilde{\epsilon}_{\pi,4}:=\sup_{P_M}\max_{c_0,c_1}\Delta(P_{M|C=c_0},P_{M|C=c_1})$. Then, we have $\tilde{\epsilon}_{\pi,4}=\epsilon_{\pi,4}$.
\end{hodai}
\begin{IEEEproof}
The proof is shown in the same way as that of $\epsilon_{\pi,3}=\epsilon_{\pi,5}$ in Lemma \ref{TS/IND}. 
\end{IEEEproof}

\begin{hodai}\label{Lem_gap2}
For any symmetric-key encryption $\pi$, it holds that 
\begin{eqnarray*}
\tilde{\epsilon}_{\pi,4}\le 2 \epsilon_{\pi,3}  \left( \sup_{P_M} P_C(c_{\rm min})^{-1} \right),  
\end{eqnarray*}
where $\displaystyle c_{\rm min}:=\arg \min_{c \in {\rm Supp}(P_C)} P_C(c)$. 
\end{hodai}
\begin{IEEEproof}
For any $P_M$, $m \in {\rm Supp}(P_M)$, and $c \in {\rm Supp}(P_C)$, we have $|P_C(c)-P_{C|M}(c|m)| \le \epsilon_{\pi,3}$, which is equivalent to 
\begin{eqnarray}
1-\frac{\epsilon_{\pi,3}}{P_C(c)} \le \frac{P_{C|M}(c|m)}{P_C(c)} \le 1+ \frac{\epsilon_{\pi,3}}{P_C(c)}. \label{NE_siki1} 
\end{eqnarray}

For any $P_M$, and $c_0,c_1\in {\rm Supp}(P_C)$, it holds that 
\begin{eqnarray}
\Delta (P_{M|C=c_0},P_{M|C=c_1})&=&\sum_{m}| P_{M|C}(m|c_0)- P_{M|C}(m|c_1) | \nonumber \\
&=&\sum_{m} P_M(m) \left|  \frac{P_{C|M}(c_0|m)}{P_{C}(c_0)} - \frac{P_{C|M}(c_1|m)}{P_{C}(c_1)} \right| \nonumber \\
&\le& \sum_{m} P_M(m) \max_{c \in {\rm Supp}(P_C)} \frac{2\epsilon_{\pi,3}}{P_C(c)}  \label{NE_siki2} \\
&=&\max_{c \in {\rm Supp}(P_C)} \frac{2\epsilon_{\pi,3}}{P_C(c)} \nonumber \\
&=&\frac{2\epsilon_{\pi,3}}{P_C(c_{\rm min})}, 
\end{eqnarray}
where $\displaystyle c_{\rm min}:=\arg \min_{c \in {\rm Supp}(P_C)} P_C(c)$ and the inequality (\ref{NE_siki2}) follows from (\ref{NE_siki1}).  
By taking the supremum over $P_M\in {\cal P}({\cal M})$, the inequality in the lemma is induced. 
\end{IEEEproof}

\medskip

\begin{meidai}\label{C_prop1}
For a symmetric-key encryption $\pi$, it holds that  
\begin{eqnarray*}
{\epsilon}_{\pi,4}\le \frac{ 2 \epsilon_{\pi,3} \cdot |{\cal C}|}{1-\sqrt{2\ln 2} \left[  
 \log |{\cal C}| -(1-\delta_{\pi,1})\log |{\cal M}| + h(\delta_{\pi,1}) \right]^{\frac{1}{2}}}.
\end{eqnarray*}
\end{meidai}
\begin{IEEEproof} 
First, we have the inequality 
\begin{eqnarray}
&&H(M) = H(M \mid K) \le H(M,C \mid K) = H(C \mid K) + H( M \mid K,C) \nonumber \\
&&\le H(C) + H(M \mid \tilde{M}) \nonumber \\
&&\le H(C) + \Pr \{ M\not=\tilde{M} \} \log (|{\cal M}|-1) + h(\Pr \{ M\not=\tilde{M}\} ) \label{NE_siki3} \\
&&\le H(C) + \delta_{\pi,1} \log |{\cal M}| + h(\delta_{\pi,1}), \label{NE_siki4}
\end{eqnarray}
where the inequality (\ref{NE_siki3}) follows from Fano's inequality.  

For the case of uniform distribution $P_M$ over ${\cal M}$, we have 
\begin{eqnarray}
D(P_{C} \parallel P_{U})&=&\log |{\cal C}| - H(C) \nonumber \\
&\le & \log |{\cal C}| - H(M) + \delta_{\pi,1} \log |{\cal M}| + h(\delta_{\pi,1}) \label{NE_siki5} \\
&=& \log |{\cal C}| - \log |{\cal M}| + \delta_{\pi,1} \log |{\cal M}| + h(\delta_{\pi,1}) \nonumber \\
&=& \log |{\cal C}| - (1-\delta_{\pi,1}) \log |{\cal M}| + h(\delta_{\pi,1}), \label{NE_siki6}
\end{eqnarray}
where (\ref{NE_siki5}) follows from (\ref{NE_siki4}). Let $\eta :=\log |{\cal C}| - (1-\delta_{\pi,1}) \log |{\cal M}| + h(\delta_{\pi,1}) $. Then, we have 
\begin{eqnarray*}
\Delta (P_C, P_U)\le \sqrt{\frac{\ln 2}{2}} D(P_{C} \parallel P_{U})^{\frac{1}{2}}\le \sqrt{\frac{\ln 2}{2}} \eta^{\frac{1}{2}}, 
\end{eqnarray*} 
and hence, we get  
\begin{eqnarray*}
 \sqrt{\frac{\ln 2}{2}} \eta^{\frac{1}{2}} &\ge& \frac{1}{2} \sum_{c} \left| P_C(c)-\frac{1}{|{\cal C}|} \right| \\
 &\ge& \frac{1}{2} |{\cal C}| \left( \frac{1}{|{\cal C}|} - P_C(c_{\rm min}) \right).
\end{eqnarray*}
Therefore, we have 
\begin{eqnarray*}
P_C(c_{\rm min})^{-1}\le \frac{|{\cal C}|}{1-\sqrt{2\ln 2}\ \eta^{\frac{1}{2}}}.  
\end{eqnarray*}
From the above inequality and Lemmas \ref{Lem_gap1} and \ref{Lem_gap2}, it follows that 
\begin{eqnarray*}
\epsilon_{\pi,4}&=&\tilde{\epsilon}_{\pi,4}\\
&\le &2 \epsilon_{\pi,3}  \left( \sup_{P_M} P_C(c_{\rm min})^{-1} \right)\\
&\le&\frac{ 2 \epsilon_{\pi,3} \cdot |{\cal C}|}{1-\sqrt{2\ln 2}\ \eta^{\frac{1}{2}}}\\
&=& \frac{ 2 \epsilon_{\pi,3} \cdot |{\cal C}|}{1-\sqrt{2\ln 2} \left[  
 \log |{\cal C}| -(1-\delta_{\pi,1})\log |{\cal M}| + h(\delta_{\pi,1}) \right]^{\frac{1}{2}}}. 
\end{eqnarray*}
\end{IEEEproof}

\medskip

We are back to the proof of Theorem \ref{main2}. 
From the assumptions, $|{\cal C}|=\Theta (|{\cal M}|)$ and $\delta_{\pi,1}=o(1/\log |{\cal M}|)$, and Proposition \ref{C_prop1}, it follows that $\epsilon_{\pi,4}=O (\epsilon_{\pi,3}\cdot |{\cal M}|)$. Here, we note that $\epsilon_{\pi,3}=\Theta (\epsilon_{\pi,5})$ by Theorem \ref{main1}. Therefore, by the assumption of $\epsilon_{\pi,5} =o(1/|{\cal M}|)$, we have $\epsilon_{\pi,4} =o(1)$. 
\end{IEEEproof}

\subsection{Non-equivalence} 
Let $\pi$ be a symmetric key encryption. 
We denote by $\mathbb{P}_{C|M}^{\pi}$ an $|\c{C}| \times |\c{M}|$ transition probability matrix associated with $\{P_{C|M}(c|m)\}_{c\in\c{C},m\in\c{M}}$ of $\pi$, i.e., each entry of $\mathbb{P}_{C|M}^{\pi}$ corresponds to $P_{C|M}(c|m)$ for $c \in \c{C}$ and $m \in \c{M}$ in $\pi$. 

The following theorem states the property of $\mathbb{P}_{C|M}^{\pi}$ for a symmetric key encryption $\pi$.

\medskip

\begin{teiri}\label{thm:fundamental}
For any symmetric key encryption $\pi$ satisfying $|\c{C}|=|\c{M}|$, its probability transition  matrix $\mathbb{P}_{C|M}^{\pi}$ is doubly stochastic\footnote{An $n \times n$ probability transition matrix $\bb{P}=(p_{i,j})$ is said to be {\em doubly stochastic} if $\sum_{i}p_{i,j}=\sum_{j}p_{i,j}=1$ for every $1\le i,j\le n$.}. 
Conversely, for any $n\times n$ matrix $A$ which is doubly stochastic, there exists a symmetric-key encryption $\pi$ such that $|\c{C}|=|\c{M}|=n$ and $\bb{P}_{C|M}^{\pi}=A$.   
\end{teiri}
\begin{IEEEproof}
In what follows, we consider the case of $|\c{M}|=|\c{C}|$. In this case, if $k \in \c{K}$ is fixed, there exists a bijection $f_k: \c{M}\rightarrow \c{C}$ since every ciphertext $c\in \c{C}$ can be uniquely decrypted by $k \in \c{K}$. Hence, for each $k \in \c{K}$, let $F_k \in \{0,1\}^{n \times n}$ be a permutation matrix which corresponds to the bijection $f_k$. Then, it is easy to see that the probability transition matrix can be represented as 
\begin{eqnarray}\label{eq:decomposition}
\bb{P}_{C|M}^{\pi} = \sum_{k\in \c{K}}P_K(k) F_k,
\end{eqnarray}
which is doubly stochastic. Conversely, due to Birknoff--von Neumann Theorem, there exists a pair of $P_K(k)$ and $F_k$, $k \in \c{K}$, satisfying \eqref{eq:decomposition}  if $\bb{P}_{C|M}^{\pi}$ is doubly stochastic. 
\end{IEEEproof}

\medskip

In the following, let $\bar{\Pi}:=\{ \pi \in \Pi \mid |{\cal C}|=\Theta (|{\cal M}|) \mbox{ and } \delta_{\pi,1}=o(1/ \log |{\cal M}|)  \}$. The following theorems show the explicit conditions for non-equivalence between security formalizations.  

\medskip

\begin{teiri}\label{gap_Th1} 
For arbitrary $\pi\in \bar{\Pi}$, if security of Type $(i,1)$ and Type $(i,5)$ are not equivalent (i.e., security of Type $(i,1)$ is asymptotically stronger than that of Type $(i,5)$) for $1\le i\le 3$, we have $\epsilon_{\pi,5} =\Omega(1/ \log |{\cal M}|)$. Conversely, for arbitrarily given $\epsilon$ such that $\epsilon=o(1)$ and $\epsilon=\Omega(1/\log n)$, there exists a symmetric-key encryption $\pi\in \bar{\Pi}$ such that, $\epsilon_{\pi,5}=\epsilon$, $n=|{\cal M}|$, and $\pi$'s security of Type $(i,1)$ and Type $(i,5)$ are not equivalent for $1\le i\le 3$.
\end{teiri}
\begin{IEEEproof} 
We show the first statement of Theorem \ref{gap_Th1} by its contraposition, namely, we prove that $\epsilon_{\pi,1}=o(1)$ for arbitrarily given $\pi\in \bar{\Pi}$ satisfying $\epsilon_{\pi,5} =o(1/\log n)$. 
This statement directly follows from (ii) of Theorem \ref{main2}. 

What remains to be shown is the second statement of Theorem \ref{gap_Th1}, and it is sufficient to prove the following proposition. 
\end{IEEEproof}

\medskip

\begin{meidai}\label{proof_gap1_prop1}
Suppose that $\epsilon=o(1)$ and $\epsilon =\Omega (1/\log n)$.  
Then, there exists a symmetric-key encryption $\pi\in \bar{\Pi}$ such that, $\epsilon_{\pi,5}=\epsilon$, $n=|{\cal M}|$, and $\pi$'s security of Type $(i,1)$ and Type $(i,5)$ are not equivalent for $1\le i\le 3$.
\end{meidai}
\begin{IEEEproof} 
For arbitarary $\epsilon$ and any positive integer $n$, we consider an  
$n \times n$ matrix $A=(a_{ij})$ defined by 
\begin{eqnarray} \label{proof_gap1_siki2}
a_{ij}=\left\{
  \begin{array}{l}
    \epsilon +\frac{1-\epsilon}{n} \quad \mbox{if $i=j$}, \\
    \frac{1-\epsilon}{n} \qquad \ \ \mbox{otherwise}. \\
  \end{array}
\right.
\end{eqnarray}
Then, it holds that $\sum_{j}a_{ij}=1$ for every $i$, and $\sum_{i}a_{ij}=1$ for every $j$, which shows that $A$ is doubly stochastic. 
Therefore, by Theorem \ref{thm:fundamental}, it follows that there exists a symmetric-key encryption $\pi$ such that $|{\cal M}|=|{\cal C}|=n$ and $\bb{P}_{C|M}^{\pi}=A$, where $\bb{P}_{C|M}^{\pi}$ is the probability transition matrix of $\pi$. Suppose that ${\cal M}=\{m_1,m_2,\ldots,m_n \}$, ${\cal C}=\{c_1,c_2,\ldots,c_n \}$, and the $(i,j)$-entry of $\bb{P}_{C|M}^{\pi}$ is equal to $P_{C|M}(c_i|m_j)$. 

It is easy to see that $\Delta(P_{C|M=m_i},P_{C|M=m_j})=\epsilon$ for every pair of $m_i,m_j \in \c{M}$ with $m_i\not=m_j$. Hence, we have $\epsilon_{\pi,5}= \epsilon$, and by taking $\epsilon \to 0$ it holds that 
\begin{eqnarray}
\lim_{\epsilon \to 0} \epsilon_{\pi,5}=0. \label{NE_siki7}
\end{eqnarray}  

On the other hand, if we assume that $P_M$ is the uniform distribution, it holds that by direct calculation  
\begin{eqnarray}
I(M;C)=\left( \epsilon + \frac{1}{n} \right) \log n 
+ \left( \epsilon + \frac{1-\epsilon}{n} \right) \log \left( \epsilon + \frac{1-\epsilon}{n} \right) 
+ \left( 1- \frac{1}{n} \right) \left( 1- \epsilon \right) \log \left( 1- \epsilon \right)
-\frac{\epsilon}{n}\log n. \label{proof_gap1_siki3}
\end{eqnarray}
Thus, by setting $\epsilon=\frac{1}{\log n}$ and taking $n\to \infty $ in (\ref{proof_gap1_siki3}), we have  
$\displaystyle \lim_{n\to \infty} I(M;C)=1$, and hence, 
\begin{eqnarray}
\lim_{\epsilon \to 0} \epsilon_{\pi,1} \ge 1. \label{NE_siki8}
\end{eqnarray}

Therefore, the proof is completed by (\ref{NE_siki7}) and (\ref{NE_siki8}).
\end{IEEEproof}

\medskip

\begin{teiri}\label{gap_Th2} 
For arbitrary $\pi\in \bar{\Pi}$, if security of Type $(i,4)$ and Type $(i,5)$ are not equivalent (i.e., security of Type $(i,4)$ is asymptotically stronger than that of Type $(i,5)$) for $1\le i\le 3$, we have $\epsilon_{\pi,5} =\Omega(1/ |{\cal M}|)$. Conversely, for arbitrarily given $\epsilon$ such that $\epsilon=o(1)$ and $\epsilon=\Omega(1/n)$, there exists a symmetric-key encryption $\pi\in \bar{\Pi}$ such that, $\epsilon_{\pi,5}=\epsilon$, $n=|{\cal M}|$, and $\pi$'s security of Type $(i,4)$ and Type $(i,5)$ are not equivalent for $1\le i\le 3$. 
\end{teiri}
\begin{IEEEproof} 
The contraposition of the first statement of Theorem \ref{gap_Th2} follows from (iii) of Theorem \ref{main2}.

In order to show the second statement of Theorem \ref{gap_Th2}, it is sufficient to prove the following proposition. 
\end{IEEEproof}

\medskip

\begin{meidai}\label{proof_gap2_prop2}
Suppose that $\epsilon=o(1)$ and $\epsilon =\Omega (1/n)$.  
Then, there exists a symmetric-key encryption $\pi\in \bar{\Pi}$ such that, $\epsilon_{\pi,5}=\epsilon$, $n=|{\cal M}|$, and $\pi$'s security of Type $(i,4)$ and Type $(i,5)$ are not equivalent for $1\le i\le 3$.
\end{meidai}
\begin{IEEEproof}
We consider a symmetric-key encryption $\pi\in \bar{\Pi}$ whose probability transition matrix is defined by (\ref{proof_gap1_siki2}) in the proof of Proposition \ref{proof_gap1_prop1}. Then, we have $\epsilon_{\pi,5}= \epsilon$ and $\displaystyle \lim_{\epsilon \to 0} \epsilon_{\pi,5}=0$ by (\ref{NE_siki7}). 

On the other hand, we derive a lower bound on $\epsilon_{\pi,4}$ as follows. For any distribution $P_M$, we have 
\begin{eqnarray*}
P_{M|C}(m_i | c_j)=\left\{
  \begin{array}{l}
    \frac{P_M(m_i)[\epsilon n +(1-\epsilon)]}{P_M(m_i)\epsilon n+ (1-\epsilon)} \quad \mbox{if $i=j$},   \\
    \frac{P_M(m_i)(1-\epsilon)}{P_M(m_j)\epsilon n+ (1-\epsilon)}  \quad \mbox{if $i\not=j$}. \\
  \end{array}
\right.
\end{eqnarray*}  
Hence, for $c_{s},c_{t}\in {\cal C}$ with $s\not=t$ and $P_M(m_{s})\le P_M(m_{t})$, we have 
\begin{eqnarray*}
\Delta (P_{M|C=c_{s}}, P_{M|C=c_{t}})=
\frac{P_M(m_s)P_M(m_t)\epsilon^2 n^2+P_M(m_t)\epsilon n (1-\epsilon) (1-P_M(m_t)+P_M(m_s))}{\left[ P_M(m_s)\epsilon n +(1-\epsilon) \right] \left[ P_M(m_t)\epsilon n +(1-\epsilon) \right]}.
\end{eqnarray*}
In particular, for the case of a distribution $P_M$ with $P_M(m_{s})=P_M(m_{t})=1/3$ and $\epsilon=1/n$, it holds that 
\begin{eqnarray}
\Delta (P_{M|C=c_{s}}, P_{M|C=c_{t}})=\frac{1}{4-\frac{3}{n}}.\label{proof_gap2_siki1}
\end{eqnarray}
Thus, we have 
\begin{eqnarray}
\epsilon_{\pi,4}&=&\sup_{P_M}\max_{c_s,c_t}\Delta (P_{M|C=c_{s}}, P_{M|C=c_{t}}) \label{proof_gap2_siki2}\\
&\ge& \frac{1}{4-\frac{3}{n}}, \label{proof_gap2_siki3}
\end{eqnarray}
where (\ref{proof_gap2_siki2}) and (\ref{proof_gap2_siki3}) follows from Lemma \ref{Lem_gap1} and (\ref{proof_gap2_siki1}), respectively. Therefore, we obtain $\displaystyle \lim_{\epsilon \to 0} \epsilon_{\pi,4}\ge 1/4$. 

From the above discussion, it follows that $\displaystyle \lim_{\epsilon \to 0} \epsilon_{\pi,4}\ge 1/4$ and $\displaystyle \lim_{\epsilon \to 0} \epsilon_{\pi,5}=0$, and the proof is completed. 
\end{IEEEproof}

\subsection{Lower Bounds and Impossibility Results}\label{sec_bound_onetime}
In this section, under each of the security formalizations in Definition \ref{enc_def_security}, we derive lower bounds on the adversary's (or distinguisher's) advantage and the required size of secret-keys. First, we note the following lower bound shown in \cite{POPE08}.  

\medskip

\begin{meidai}[\cite{POPE08}] \label{bound101}
Let $\pi$ be a symmetric-key encryption protocol. 
Then, for any simulator $\sigma$ on ${\cal C}$, and for the set of all distinguishers ${\cal D}$, we have 
\begin{eqnarray*}
 \Delta^{{\cal D}}(\pi (\LAC \ ||\ [\mbox{$P_{K}$}]),\sigma (\SEC \ ))
 \ge 1-\frac{|{\cal K}|}{|{\cal M}|}.  \label{ineq2}
\end{eqnarray*}
\end{meidai}

\medskip

In \cite{POPE08} Pope showed the above lower bound by only considering a distinguisher that inputs the uniformly distributed plaintexts into the symmetric-key encryption protocol for distinguishing real output and ideal one. From the above proposition, it follows that 
\begin{eqnarray}
\epsilon_{\pi,8}\ge 1-\frac{|{\cal K}|}{|{\cal M}|}, \label{LB_enc_siki1}
\end{eqnarray}
for arbitrary symmetric-key encryption $\pi$ and [$P_K$].
We now derive lower bounds for the adversary's (or distinguisher's) advantage under all formalizations in Definition \ref{enc_def_security} at once through our relationships.   

\medskip

\begin{teiri}\label{bound103}
For any symmetric-key encryption protocol $\pi$ and [$P_K$], we have:  
\begin{eqnarray*}
&&\mbox{(i)}\quad  \delta_{\pi,i} + \epsilon_{\pi,j} \ge 1-\frac{|{\cal K}|}{|{\cal M}|} \quad \mbox{for } i\in \{1,2,3 \} \mbox{ and } j\in \{ 3,5,6,8,9,10 \}, \\
&&\mbox{(ii)}\quad \delta_{\pi,i} + 2\epsilon_{\pi,j} \ge 1-\frac{|{\cal K}|}{|{\cal M}|} \quad \mbox{for } i\in \{1,2,3 \} \mbox{ and } j\in \{ 2,4 \},\\
&&\mbox{(iii)}\quad \delta_{\pi,i} + 4\epsilon_{\pi,7} \ge 1-\frac{|{\cal K}|}{|{\cal M}|} \quad \mbox{for } i\in \{1,2,3 \},\\
&&\mbox{(iv)}\quad \delta_{\pi,i} + \sqrt{\frac{\ln 2}{2}}\epsilon_{\pi,1}^{\frac{1}{2}} \ge 1-\frac{|{\cal K}|}{|{\cal M}|} \mbox{ for }  i\in \{1,2,3 \}, 
\end{eqnarray*}
where $\delta_{\pi,i}$ and $\epsilon_{\pi,j}$ are parameters of formalizations of correctness and security, respectively, defined in Definition \ref{enc_def_security}.
\end{teiri}
\begin{IEEEproof}
By Theorem \ref{main1}, we have 
\begin{eqnarray*}
&&\epsilon_{\pi,8}\le \epsilon_{\pi,9}+\delta_{\pi,2},\\
&&\delta_{\pi,1}=\delta_{\pi,2}=\delta_{\pi,3}, \\
&&\epsilon_{\pi,9}=\epsilon_{\pi,10}\le \epsilon_{\pi,3}=\epsilon_{\pi,5}=\epsilon_{\pi,6}.
\end{eqnarray*}
Combining the above inequalities with (\ref{LB_enc_siki1}), we obtain (i). 

In addition, by Theorem \ref{main1}, we have $\epsilon_{\pi,3}\le 2\epsilon_{\pi,2}$ and $\epsilon_{\pi,2}\le \epsilon_{\pi,4}$. Therefore, we have (ii) by these inequalities and (i).  

Similarly, the inequalities $\epsilon_{\pi,6}\le 4\epsilon_{\pi,7}$ and $\epsilon_{\pi,2}\le \sqrt{\frac{\ln 2}{2}} \epsilon_{\pi,1}^{1/2}$ shown by Theorem \ref{main1} imply (iii) and (iv), respectively, by applying them to (i). 
\end{IEEEproof}

\medskip

From Theorem \ref{bound103}, we obtain the following lower bounds on the size of secret-keys. The proof immediately follows from Theorem \ref{bound103}, and we omit the proof. 

\medskip

\begin{kei}\label{bound104}
Suppose that a symmetric-key encryption protocol $\pi$ is $(\delta,\epsilon)$-secure in the sense of Type $(i,j)$. Then, we have the following lower bounds on the size of secret-keys:  
\begin{eqnarray*}
&&\mbox{(i)}\quad |{\cal K}| \ge \left\{ 1-  \left( \delta + \epsilon \right) \right\} |{\cal M}| \quad \mbox{for $i\in \{1,2,3 \}$ and $j\in \{3,5,6,8,9,10 \}$}, \\
&&\mbox{(ii)}\quad  |{\cal K}| \ge \left\{ 1-  \left( \delta + 2\epsilon \right) \right\} |{\cal M}| \quad \mbox{for $i\in \{1,2,3 \}$ and $j\in \{ 2,4 \}$}, \\
&&\mbox{(iii)}\quad  |{\cal K}| \ge \left\{ 1-  \left( \delta + 4\epsilon \right) \right\} |{\cal M}| \quad \mbox{for $i\in \{1,2,3 \}$ and $j=7$}, \\
&&\mbox{(iv)}\quad  |{\cal K}| \ge \left\{ 1-  \left( \delta + \sqrt{2\ln 2}\epsilon^{\frac{1}{2}} \right) \right\}|{\cal M}| \quad \mbox{for $i\in \{1,2,3 \}$ and $j=1$}.
\end{eqnarray*}
\end{kei}

\medskip

\begin{tyuui} {\rm 
As described in \cite{Shoup}, it is known that: {\it Let $\{ \Phi_r | r\in {\cal R} \}$ be a family of (hash) functions from ${\cal S}$ to ${\cal T}$ such that: each $\Phi_r$ maps ${\cal S}$ injectively into ${\cal T}$; and there exists $\epsilon \in [0,1]$ such that $\Delta ( \Phi_H(s), \Phi_H(s') ) \le \epsilon$ for all $s,s'\in {\cal S}$, where $H$ is uniformly distributed over ${\cal R}$. Then, we have $|{\cal R}|\ge (1-\epsilon)|{\cal S}|$.}   
Corollary \ref{bound104} can be understood as an extension of the above statement (see (i) in Corollary \ref{bound104}). Actually, we do not necessarily assume that: $P_K$ is uniform; or for each $k\in {\cal K}$, $\pi^A(k,\cdot):{\cal M}\to {\cal C}$ is deterministic and injective (Note that $\delta$ can be zero if $\pi^A(k,\cdot)$ is injective). 
}
\end{tyuui}

\medskip

\begin{tyuui} 
{\rm
In \cite{Dodis12}, Dodis derives the lower bound (i) in Corollary \ref{bound104}, and shows that this bound is tight with respect to $\delta$ and $\epsilon$ up to a constant. In fact, by using a mechanism of the one-time pad, two constructions satisfying the following parameters are proposed in \cite{Dodis12}: $\epsilon=0$ and $|{\cal K}|=(1-\delta)|{\cal M}|$ for given $\delta\in [0,1]$; and $\delta=0$ and $|{\cal K}|=(1-\frac{1}{2}\epsilon)|{\cal M}|$ for given $\epsilon\in [0,1]$ such that ${\epsilon}\cdot |{\cal M}| /2$ is non-negative integer. By the constructions, it is straightforwardly seen that our lower bounds in Corollary \ref{bound104} are also tight with respect to $\delta$ and $\epsilon$ up to a constant.  
}
\end{tyuui}

\medskip

By considering a contraposition of Corollary \ref{bound104}, we obtain the following impossibility result: There exists no symmetric-key encryption protocol which is $(\delta,\epsilon)$-secure in the sense of Type $(i,j)$, if $\delta$ and $\epsilon$ are some real numbers such that they do not satisfy the corresponding inequality among (i)--(iv) in Corollary \ref{bound104}. 

\section{Key Agreement}
\subsection{Protocol Execution} \label{PE-KA} 
We explain protocol execution of key agreement. 
Let ${\cal X}$ and ${\cal Y}$ be finite sets. Suppose that Alice and Bob can have access to an ideal resource, and that they can finally obtain $x\in {\cal X}$ and $y\in {\cal Y}$, respectively. For simplicity, suppose that the resource is given by a correlated randomness resource [$P_{XY}$].    
In addition, we assume that there is the bidirectional (or unidirectional) authenticated channel available between Alice and Bob, and that Eve can eavesdrop on all information transmitted by the channel without any error.  
    
Let ${\cal K}$ be a set of keys, and let $K$ be a random variable which takes values on ${\cal K}$ in $\KA $ (or more generally, [$P_K$]). 
Also, let ${\cal T}$ be a set of transcripts between Alice and Bob.    
Let $\pi=(\pi^{A}_{ka},\pi^{B}_{ka})$ be a key agreement protocol, where $\pi^{A}_{ka}$ (resp. $\pi^{B}_{ka}$) is a converter at Alice's (resp. Bob's) side, defined below: Let $l$ be a positive integer and $\lambda=2l-1$; The converter $\pi^{A}_{ka}$ consists of (probabilistic) functions $f_1,f_3,f_5,\ldots,f_{2l-1}$ and $g_A$, and the converter $\pi^{B}_{ka}$ consists of (probabilistic) functions $f_2,f_4,f_6,\ldots,f_{2l-2}$ and $g_B$, where the functions $f_1,f_2,\ldots,f_{n},g_A,g_B$ are defined as follows: 
\begin{eqnarray*}
&&f_i: {\cal X}\times {\cal T}^{i-1}\to {\cal T}, \ t_i=f_i(x,t_1,t_2,t_3,\ldots,t_{i-1}) \mbox{ for $i=1,3,\ldots,2l-1$};\\
&&f_j: {\cal Y}\times {\cal T}^{j-1}\to {\cal T}, \ t_j=f_j(y,t_1,t_2,t_3,\ldots,t_{j-1}) \mbox{ for $j=2,4,\ldots,2l-2$};\\
&&g_A: {\cal X}\times {\cal T}^{\lambda}\to {\cal K}, \ k_A=g_{A}(x,t_1,t_2,t_3,\ldots,t_{\lambda}); \ \ 
g_B: {\cal Y}\times {\cal T}^{\lambda}\to {\cal K}, \ k_B=g_B(y,t_1,t_2,t_3,\ldots,t_{\lambda}).
\end{eqnarray*}

\begin{table}[h]
 \begin{center}
  \begin{tabular}{l}
       \hline
      {\bf Key Agreement Protocol $\pi$}\\
       \hline \hline
      {Input of Alice's inner interface}: $x\in {\cal X}$ by accessing [$P_{XY}$] \\
      {Input of Bob's inner interface}: $y\in {\cal Y}$ by accessing [$P_{XY}$] \\
      {Output of Alice's outer interface}: $k_A\in {\cal K}$ \\ 
      {Output of Bob's outer interface}: $k_B\in {\cal K}$ \\
      1. $\pi^A_{ka}$ computes $t_{1}=f_{1}(x)$ and sends $t_{1}$ to $\pi^B_{ka}$ by $\LAC\ $.\\
      2. For $s$ from $1$ to $(\lambda-1)/2$,      \\
      \qquad 2-1. $\pi^B_{ka}$ computes $t_{2s}=f_{2s}(y,t_1,t_2,\ldots,t_{2s-1})$. Then, $\pi^B_{ka}$ sends $t_{2s}$ to $\pi^A_{ka}$ by $\RAC\ $. \\ 
      \qquad 2-2. $\pi^A_{ka}$ computes $t_{2s+1}=f_{2s+1}(x,t_1,t_2,\ldots,t_{2s})$. Then, $\pi^A_{ka}$ sends $t_{2s+1}$ to $\pi^B_{ka}$ by $\LAC\ $. \\
           3. $\pi^A_{ka}$ computes $k_A=g_{A}(x,t_1,t_2,\ldots,t_{\lambda})$ and outputs $k_A$. \\
      \ \ \ Similarly, $\pi^B_{ka}$ computes $k_B=g_{B}(y,t_1,t_2,\ldots,t_{\lambda})$ and outputs $k_B$.   \\
    \hline
  \end{tabular}
 \end{center}
\end{table}
\vspace*{-5mm}

Note that, if only the unidirectional authenticated channel from Alice to Bob is available, the functions $f_i$ for even $i$ could be understood as trivial functions which always return a certain single point (or symbol). Similarly, we can capture the case of only the unidirectional authenticated channel from Bob to Alice being available.

For every $i$ with $1\le i\le \lambda$, $T_i$ denotes a random variable which takes values $t_i\in {\cal T}$, and let $T^{\lambda}:=(T_1,T_2,\ldots,T_{\lambda})$ be the joint random variable which takes values $t^{\lambda}=(t_1,t_2,\ldots,t_{\lambda})\in {\cal T}^{\lambda}$. 
Also, let $K_A$ and $K_B$ be the random variables which take values $k_A\in {\cal K}$ and $k_B\in {\cal K}$, respectively. 

For simplicity, we assume that a key agreement protocol $\pi$ can be used at most one time (i.e., we deal with key agreement protocols in the one-time model). Therefore, the purpose of the key agreement protocol is to transform a correlated randomness resource [$P_{XY}$] and channels $(\LAC\ )^l \parallel (\RAC\ )^{l-1}$ into a key sharing resource $\KA $ ( or more generally, [$P_K$]). 
\subsection{Security Definitions Revisited: Formalizations and Relationships} 
As in the case of symmetric-key encryption protocols, let's consider the following traditional formalization of security for key agreement protocols (e.g. \cite{Csis96, CN00, DM04, Maurer93, Maurer94, MW03}).

\medskip

\begin{teigi}{\rm  \label{sta_ka_def}
Let $\pi$ be a key agreement protocol. Then, $\pi$ is said to be {\it $\epsilon$-secure} if it satisfies the following conditions: 
\begin{eqnarray*}
\Pr \{ K_A \neq K_B\} \le \epsilon, \ \log |{\cal K}|-H(K_A) \le \epsilon, \mbox{ and } I(K_A; T^{\lambda})\le \epsilon.  
\end{eqnarray*}
In particular, $\pi$ is said to be {\it perfectly-secure} if $\epsilon=0$ above. 
}
\end{teigi}

\medskip

We now consider the following formalizations of information-theoretic security for key agreement. 

\medskip

\begin{teigi} \label{def_ka1}
{\rm 
Let $\pi$ be a key agreement protocol such that $P_K$ is the uniform distribution over ${\cal K}$ (i.e., [$P_K$]=$\KA $). 
We define the following formalizations of correctness and security.
\begin{itemize}
\item Correctness. We define the following parameters concerning correctness of $\pi$:  
\begin{eqnarray*}
\delta_{\pi,1}&:=&\max \{ \Pr \{ K_A \neq K_B\}, \ \log |{\cal K}|-H(K_A)\}, \\
\delta_{\pi,2}&:=&\Delta (P_{K_AK_B},P_{KK}).
\end{eqnarray*}
\item Security. We define the following advantage of adversaries for security: 
\begin{eqnarray*}
\epsilon_{\pi,1}&:=& I(K_A;T^{\lambda}), \\ 
\epsilon_{\pi,2}&:=&\Delta (P_{K_AT^{\lambda}},P_{K_A}P_{T^{\lambda}}), \\ 
\epsilon_{\pi,3}&:=&\inf_{P_Q} \Delta (P_{K_AT^{\lambda}},P_{K_A}P_Q),
\end{eqnarray*}  
where the infimum ranges over all $P_Q\in {\cal P}({\cal T}^{\lambda})$. 
\end{itemize}

Then, $\pi$ is said to be {\it $(\delta,\epsilon)$-secure in the sense of Type $(i,j)$} for $1\le i\le 2$ and $1\le j\le 3$, if $\pi$ satisfies 
$\delta_{\pi,i}\le \delta$  and  $\epsilon_{\pi,j}\le \epsilon$. 
}
\end{teigi}

\medskip

The traditional definition in Definition \ref{sta_ka_def} corresponds to the security in the sense of Type $(1,1)$. The  composable security by Maurer et al. \cite{MR11,MT10} and Canetti \cite{C01,C05} is closely related to the security in the sense of Type $(2,3)$: $\delta_{\pi,2}$ means distinguisher's advantage for distinguishing real output and ideal one at honest players' interfaces, and $\delta_{\pi,2}$ is the same as the formalization of availability in Definition \ref{sim_def} for key agreement;  $\epsilon_{\pi,3}$ means distinguisher's advantage for distinguishing real transcripts and simulator's output at $E$-interface, together with output at $A$-interface. Note that the formalization $\epsilon_{\pi,3}$ is simple, and validity of $\epsilon_{\pi,3}$ is well explained by the following proposition. 

\medskip

\begin{meidai}\label{justification2}
The formalization of security in Definition \ref{sim_def} for a key agreement protocol $\pi$ is lower-and-upper bounded as follows: 
\begin{eqnarray*}
\max \left( \frac{1}{3}\epsilon_{\pi,3},\delta_{\pi,2}\right) \le \inf_{\sigma} \Delta^{\cal D}(\pi ( (\LAC \ )^l \| (\RAC )^{l-1} \| \ [\mbox{$P_{XY}$}]),\sigma (\KA ))  
\le \epsilon_{\pi,3}+2 \delta_{\pi,2}. 
\end{eqnarray*}
\end{meidai}
\begin{IEEEproof} 
By focusing on distributions of output at $A$'s, $B$'s and $E$'s interfaces, for simplicity, we write \\ 
$\inf_{P_Q} \Delta (P_{K_A K_B T^{\lambda}},P_{KK}P_Q)$ for $\inf_{\sigma} \Delta^{\cal D}(\pi ( (\LAC \ )^l \| (\RAC )^{l-1} \| \ [\mbox{$P_{XY}$}]),\sigma (\KA ))$, where $P_{K}$ is the uniform distribution over ${\cal K}$.  

For any distribution $P_Q\in {\cal P}({\cal C})$, we have 
\begin{eqnarray*}
\Delta (P_{K_A K_B T^{\lambda}},P_{KK}P_Q) 
&\le& \Delta (P_{K_A K_B T^{\lambda}}, P_{K_A K_A T^{\lambda}})+ \Delta (P_{K_A K_A T^{\lambda}}, P_{K_AK_A}P_Q)\\
&& \  + \Delta (P_{K_AK_A}P_Q, P_{KK}P_Q)\\
&=& \Pr \{ K_A \not= K_B \} + \Delta (P_{K_A T^{\lambda}}, P_{K_A}P_Q)+\Delta (P_{K_A}, P_{K})\\
&\le &\Delta (P_{K_A T^{\lambda}}, P_{K_A}P_Q)+2 \Delta (P_{K_AK_B}, P_{KK}). 
\end{eqnarray*} 
By taking the infimum over all $P_Q\in {\cal P}({\cal T}^{\lambda})$, we have 
\begin{eqnarray*}
\inf_{P_Q} \Delta (P_{K_A K_B T^{\lambda}},P_{KK}P_Q)
&\le & \inf_{P_Q}\Delta (P_{K_A T^{\lambda}}, P_{K_A}P_Q)+2 \Delta (P_{K_AK_B}, P_{KK})\\
&=& \epsilon_{\pi,3}+2\delta_{\pi,2}. 
\end{eqnarray*} 

In addition, for any distribution $P_Q\in {\cal P}({\cal C})$ we have 
\begin{eqnarray*}
\Delta (P_{K_A T^{\lambda}}, P_{K_A}P_Q) 
&\le&  \Delta (P_{K_A K_A T^{\lambda}}, P_{K_A K_B T^{\lambda}})+ \Delta (P_{K_A K_B T^{\lambda}}, P_{KK}P_Q)+ \Delta (P_{KK}P_Q, P_{K_AK_A}P_Q)\\
&=& \Pr \{ K_A \not= K_B\}+ \Delta (P_{K_A K_B T^{\lambda}}, P_{KK}P_Q)+ \Delta (P_{K}, P_{K_A})\\
&\le& 2 \Delta (P_{K_AK_B}, P_{KK})+\Delta (P_{K_A K_B T^{\lambda}}, P_{KK}P_Q)\\
&\le& 3  \Delta (P_{K_A K_B T^{\lambda}}, P_{KK}P_Q). 
\end{eqnarray*}
By taking the infimum over all $P_Q\in {\cal P}({\cal T}^{\lambda})$, we have
\begin{eqnarray*}
\frac{1}{3}\epsilon_{\pi,3} \le \inf_{P_Q} \Delta (P_{K_A K_B T^{\lambda}},P_{KK}P_Q).  
\end{eqnarray*}

Finally, it is straightforward to see that $\delta_{\pi,2} \le \inf_{P_Q} \Delta (P_{K_A K_B T^{\lambda}}, P_{KK}P_Q)$.
\end{IEEEproof} 

\medskip
   
Then, as in the case of symmetric-key encryption, we can show the following theorem which states relationships between all the formalizations above (i.e., six possible formalizations above). 

\medskip

\begin{teiri} \label{relation_ka}
Let $\pi$ be a key agreement protocol such that $P_K$ is the uniform distribution over ${\cal K}$. 
Then, we have explicit relationships between $\delta_{\pi,i}$, $\epsilon_{\pi,j}$ for $i\in \{1,2 \}$, $j\in \{1,2,3 \}$ as follows: 
\begin{eqnarray*}
&(1)& \delta_{\pi,2}\le \delta_{\pi,1}+\sqrt{\frac{\delta_{\pi,1}\ln 2}{2}} \ \mbox{ and }\  
\delta_{\pi,1}\le -2 \delta_{\pi,2} \log \frac{2\delta_{\pi,2}}{|{\cal K}|},\\
&(2)& \frac{2}{\ln 2}\epsilon_{\pi,2}^2 \le \epsilon_{\pi,1}\le - 2\epsilon_{\pi,2} \log \frac{2\epsilon_{\pi,2}}{|{\cal K}| |{\cal T}|^{\lambda}},  \\ 
&(3)& \epsilon_{\pi,3} \le \epsilon_{\pi,2} \le 2 \epsilon_{\pi,3}.
\end{eqnarray*}

Furthermore, it holds that: 
\begin{enumerate}
\item[(i)] For arbitrary key agreement $\pi$, it holds that $\pi$'s security of Type $(i,2)$ and Type $(i,3)$ are strictly equivalent for every $i\in \{ 1,2\}$;  
\item[(ii)]  Suppose that a key agreement protocol $\pi$ satisfies $\epsilon_{\pi,2}=o(\frac{1}{\log |{\cal K}|+\lambda \log |{\cal T}|})$. Then, it holds that $\pi$'s security of Type $(i,1)$, Type $(i,2)$, and Type $(i,3)$ are equivalent for every $i\in \{ 1,2\}$;   
\item[(iii)] Suppose that a key agreement protocol $\pi$ satisfies $\delta_{\pi,2}=o(1/\log |{\cal K}|)$.  
Then, it holds that $\pi$'s security of Type $(1,j)$ and Type $(2,j)$ are equivalent for every $j\in \{ 1,2,3 \}$; 
\item[(iv)] Suppose that a key agreement protocol $\pi$ satisfies $\delta_{\pi,2}=o(1/\log |{\cal K}|)$ and $\epsilon_{\pi,2}=o(\frac{1}{\log |{\cal K}|+\lambda \log |{\cal T}|})$. Then, all $\pi$'s security of Type $(i,j)$ are equivalent.
\end{enumerate}
\end{teiri}
\begin{IEEEproof} 
First, we show (1):   
By Lemma \ref{furoku1} in Appendix B, we have 
\begin{eqnarray*}
\delta_{\pi,2}&=& \Delta (P_{K_AK_B},P_{KK})\\
&\le&\Pr \{ K_A\not= K_B\} +\min \{ \Delta(P_{K_A}, P_{K}), \Delta(P_{K_B}, P_{K})\}.
\end{eqnarray*}
In addition, by Proposition \ref{CT1} in Appendix A we have  
\begin{eqnarray*}
\Delta(P_{K_A},P_K)^2&\le& \frac{\ln 2}{2} D(P_{K_A}||P_K) \\
&=&\frac{\ln 2}{2} (\log |{\cal K}|-H(K_A))\\
&\le& \frac{\ln 2}{2} \delta_{\pi,1}. 
\end{eqnarray*}
Therefore, we have $\delta_{\pi,2}\le \delta_{\pi,1}+\sqrt{\frac{\delta_{\pi,1}\ln 2}{2}}$.  

Conversely, we have 
\begin{eqnarray}
\Pr \{ K_A \neq K_B\} &\le& \delta_{\pi,2}, \mbox{\  and } \nonumber \\
\log |{\cal K}|-H(K_A) &\le& -2 \Delta(P_{K_A},P_K) \log \frac{2\Delta(P_{K_A},P_K)}{|{\cal K}|} \label{IT1}\\
&\le& -2 \delta_{\pi,2} \log \frac{2\delta_{\pi,2}}{|{\cal K}|}, \nonumber
\end{eqnarray}
where (\ref{IT1}) follows from Proposition \ref{CT2}. Thus, we obtain 
\begin{eqnarray*}
\delta_{\pi,1}\le -2 \delta_{\pi,2} \log \frac{2\delta_{\pi,2}}{|{\cal K}|}.
\end{eqnarray*} 

Secondly, the proof of (2) is given in the same way as that of Theorem \ref{main1}, and we omit it. 

Thirdly, we show (3): By definition, we have $\epsilon_{\pi,3}\le \epsilon_{\pi,2}$. In addition, for any $\epsilon>0$, there is a distribution $P_Q$ such that $\epsilon_{\pi,3}+\epsilon \ge \Delta (P_{K_AT^{\lambda}},P_{K_A}P_Q)$. Then, we have 
\begin{eqnarray*}
\epsilon_{\pi,2} &\le& \Delta (P_{K_AT^{\lambda}},P_{K_A}P_Q)+\Delta (P_{K_A}P_Q,P_{K_A}P_{T^{\lambda}})\\
&\le& \epsilon_{\pi,3}+\epsilon +\Delta (P_Q,P_{T^{\lambda}})\\
&\le&2(\epsilon_{\pi,3}+\epsilon),  
\end{eqnarray*}
where the last inequality follows from $\Delta (P_Q,P_{T^{\lambda}})\le \Delta (P_{K_A}P_Q,P_{K_AT^{\lambda}})\le \epsilon_{\pi,3}+\epsilon$. Thus, we obtain $\epsilon_{\pi,2}\le 2 \epsilon_{\pi,3}$. 

Finally, (i) follows from (3) above; (ii) follows from Lemma \ref{proof_gap1_lem1} and (2), (3) above and ; Similarly, (iii) follows from Lemma \ref{proof_gap1_lem1} and (1) above; and (iv) follows from (ii) and (iii) above.  
\end{IEEEproof}

\subsection{Lower Bounds and Impossibility Results}
For any key agreement protocol which constructs a key sharing resource [$P_K$] starting from a correlated randomness resource [$P_{XY}$], we show a lower bound on the advantage of distinguishers as follows. The proof is given in Appendix C. 

\medskip

\begin{hodai}\label{main_ka1}
Let [$P_K$] be a key sharing resource. 
For any key agreement protocol $\pi$, and for any simulator $\sigma$, we have 
\begin{eqnarray*}
\Delta^{\cal D}(\pi ( (\LAC \ )^l \|  (\RAC )^{l-1} \|  \ [\mbox{$P_{XY}$}]),\sigma ([P_K])) \ge 1 - 2^{H_{0}(X,Y)-H_{\infty }(K)}.
\end{eqnarray*}
In particular, we have 
\begin{eqnarray*}
\Delta^{\cal D}(\pi ( (\LAC \ )^l \|  (\RAC )^{l-1} \|  \ [\mbox{$P_{XY}$}]),\sigma (\KA )) \ge 1 - \frac{2^{H_{0}(X,Y)}}{|{\cal K}|}.
\end{eqnarray*}
\end{hodai}

\medskip

From Lemma \ref{main_ka1}, we obtain lower bounds on the adversary's (or distinguisher's) advantage by Theorem \ref{bound303} below, and the required size of a correlated randomness resource by Corollary \ref{bound304} below.  

\medskip

\begin{teiri}\label{bound303}
For any key agreement protocol $\pi$ such that $P_K$ is the uniform distribution over ${\cal K}$, we have the following lower bounds on the adversary's advantage:  
\begin{eqnarray*}
&&\mbox{(i)}\quad  2 \left( 1+ \sqrt{\frac{\ln 2}{2}} \right) \delta_{\pi,1}^{\frac{1}{2}} + \sqrt{\frac{\ln 2}{2}}\epsilon_{\pi,1}^{\frac{1}{2}} \ge 1-\frac{2^{H_0(X,Y)}}{|{\cal K}|}, \mbox{ \ if  $ \delta_{\pi,1} \in [0,1]$}; \\ 
&&\mbox{(ii)}\quad 2 \left( 1+ \sqrt{\frac{\ln 2}{2}} \right) \delta_{\pi,1}^{\frac{1}{2}} + \epsilon_{\pi,j}
\ge 1-\frac{2^{H_0(X,Y)}}{|{\cal K}|}  \mbox{ \ for $j\in \{2,3 \}$, if $ \delta_{\pi,1} \in [0,1]$}; \\ 
&&\mbox{(iii)}\quad 2\delta_{\pi,2} + \sqrt{\frac{\ln 2}{2}}\epsilon_{\pi,1}^{\frac{1}{2}} 
\ge  1-\frac{2^{H_0(X,Y)}}{|{\cal K}|}; \\
&&\mbox{(iv)}\quad 2\delta_{\pi,2} + \epsilon_{\pi,j} \ge  1-\frac{2^{H_0(X,Y)}}{|{\cal K}|} \mbox{ \ for $j\in \{2,3 \}$,}
\end{eqnarray*}
where $\delta_{\pi,i}$ and $\epsilon_{\pi,j}$ are parameters of formalizations of correctness and security, respectively, defined in Definition \ref{def_ka1}.
\end{teiri}
\begin{IEEEproof} 
By Proposition \ref{justification2}, we have    
\begin{eqnarray}
\inf_{\sigma} \Delta^{\cal D}( \pi ((\LAC \ )^l \| (\RAC )^{l-1} \| \ [\mbox{$P_{XY}$}]), \sigma (\KA ) ) 
\le \epsilon_{\pi,3}+2\delta_{\pi,2}. \label{siki82} 
\end{eqnarray} 

Therefore, by (\ref{siki82}) and Lemma \ref{main_ka1} we obtain 
\begin{eqnarray*}
\epsilon_{\pi,3}+2\delta_{\pi,2}\ge 1 - \frac{2^{H_{0}(X,Y)}}{|{\cal K}|}. 
\end{eqnarray*}

By Theorem \ref{relation_ka}, we have explict relationships between $\delta_{\pi,i}$ and $\epsilon_{\pi,j}$ as follows: 
\begin{eqnarray*}
\delta_{\pi,2} &\le& \delta_{\pi,1}+ \sqrt{\frac{\ln 2}{2}}\delta_{\pi,1}^{\frac{1}{2}}\\
&\le& \left( 1+ \sqrt{\frac{\ln 2}{2}} \right) \delta_{\pi,1}^{\frac{1}{2}} \mbox{ \ if  $ \delta_{\pi,1} \in [0,1]$}; \\
\epsilon_{\pi,3}&\le& \epsilon_{\pi,2} \le \sqrt{\frac{\ln 2}{2}}\epsilon_{\pi,1}^{\frac{1}{2}}.
\end{eqnarray*}

Therefore, by combining the above inequalities we obtain all lower bounds in Theorem \ref{bound303}. 
\end{IEEEproof}

\medskip

\begin{kei}\label{bound304}
Suppose that a key agreement protocol $\pi$ is $(\delta,\epsilon)$-secure in the sense of Type $(i,j)$ in which $P_K$ is the uniform distribution over ${\cal K}$. Then, we have the following lower bounds on the size of a correlated randomness resource:  
\begin{eqnarray*}
&&\mbox{(i)}\quad  2^{H_0(X,Y)} \ge \left\{ 1-\left[ \sqrt{\frac{\ln 2}{2}}\epsilon^{\frac{1}{2}}+ 2\left( 1+ \sqrt{\frac{\ln 2}{2}} \right) \delta^{\frac{1}{2}} \right] \right\} |{\cal K}| \mbox{ \ for $i=j=1$, \ if  $ \delta \in [0,1]$}; \\ 
&&\mbox{(ii)} \quad  2^{H_0(X,Y)} \ge \left\{1- \left[ \epsilon+2\left( 1+ \sqrt{\frac{\ln 2}{2}} \right) \delta^{\frac{1}{2}} \right] \right\} |{\cal K}| \mbox{ \ for $i=1$ and $j\in \{2,3 \}$, if $ \delta \in [0,1]$}; \\ 
&&\mbox{(iii)}\quad  2^{H_0(X,Y)} \ge \left\{ 1-\left( \sqrt{\frac{\ln 2}{2}}\epsilon^{\frac{1}{2}} +2\delta \right) \right\} |{\cal K}|  \mbox{ \ for $i=2$ and $j=1$}; \\
&&\mbox{(iv)}\quad  2^{H_0(X,Y)} \ge \left\{ 1-\left( \epsilon+2\delta \right) \right\} |{\cal K}| \mbox{ \ for $i=2$ and $j\in \{2,3 \}$.}
\end{eqnarray*}
\end{kei}
\begin{IEEEproof} 
The proof of Corollary \ref{bound304} immediately follows from Theorem \ref{bound303}.
\end{IEEEproof}

\medskip

Finally, from Lemma \ref{main_ka1} we obtain Proposition \ref{cor_ka} which is an impossibility result for key agreement.  
Also, we provide Corollaries \ref{ex_first} and \ref{ex_last} below, as illustrations of impossibility results which are special cases of Proposition \ref{cor_ka}. The proofs immediately follow from Theorem \ref{bound303} and Proposition \ref{cor_ka}, and we omit them. 

\medskip

\begin{meidai}\label{cor_ka}
Let [$P_K$] be a key sharing resource, and let [$P_{XY}$] be a correlated randomness resource.  
In addition, let $\hat{\epsilon}$ be a real number such that  
$\displaystyle \hat{\epsilon} <  1-2^{H_{0}(X,Y)-H_{\infty }(K)}$. 
Then, there exists no key agreement protocol $\pi$ such that    
$\displaystyle (\LAC \ )^{\infty } \|  (\RAC )^{\infty } \|  \ [\mbox{$P_{XY}$}] \stackrel{\pi,\hat{\epsilon}}{\Longrightarrow}  [P_K]$.  
\end{meidai}

\medskip

\begin{kei}\label{ex_first} 
There is no key agreement protocol $\pi$ such that 
 \ $\displaystyle (\LAC \ )^{\infty } \|  (\RAC )^{\infty }  \stackrel{\pi,\hat{\epsilon}}{\Longrightarrow}  [P_K]$  \  
for $\hat{\epsilon} < 1- 1/2^{H_{\infty }(K)}$. 
In particular, there is no $(\delta,\epsilon)$-secure key agreement in the sense of Type $(i,j)$ which constructs $\KA$ (even with 1-bit) starting from authenticated communications, if $\delta, \epsilon \in [0,1]$ are some real numbers such that: 
\begin{eqnarray*}
&&\mbox{(i)}\quad  \sqrt{\frac{\ln 2}{2}}\epsilon^{\frac{1}{2}}+ 2\left( 1+ \sqrt{\frac{\ln 2}{2}} \right) \delta^{\frac{1}{2}} < \frac{1}{2} \mbox{ for } i=j=1; \\   
&&\mbox{(ii)}\quad \epsilon+2\left( 1+ \sqrt{\frac{\ln 2}{2}} \right) \delta^{\frac{1}{2}} < \frac{1}{2} \mbox{ for } i=1 \mbox{ and } j\in \{ 2,3 \}; \\ 
&&\mbox{(iii)}\quad \sqrt{\frac{\ln 2}{2}}\epsilon^{\frac{1}{2}} +2\delta < \frac{1}{2} \mbox{ for } i=2 \mbox{ and } j=1; \\
&&\mbox{(iv)}\quad \epsilon+2\delta < \frac{1}{2} \mbox{ for } i=2 \mbox{ and } j\in \{2,3 \}.
\end{eqnarray*}
\end{kei} 

\medskip

\begin{kei}\label{ex_last}
Let $l$ and $s$ be nonnegative integers with $l<s$. In addition, we denote the $l$-bit key sharing resource by $\KA_l$, and let [$P_K$]$_s$ be an $s$-bit key sharing resource with min-entropy $H_{\infty}(K)$. Then, there is no protocol $\pi$ such that  \ \ 
$\displaystyle (\LAC \ )^{\infty } \|  (\RAC )^{\infty } \|  \KA_l \stackrel{\pi,\hat{\epsilon}}{\Longrightarrow}  [P_K]_s$ \ \  
for $ \hat{\epsilon} < 1-2^{l-H_{\infty}(K)}$. 
In particular, there is no $(\delta,\epsilon)$-secure key agreement (or key-expansion) protocol in the sense of Type $(i,j)$  which constructs the $s$-bit key sharing resource $\KA_s$ from the $l$-bit key sharing resource $\KA_l$, if $\delta, \epsilon \in [0,1]$ are some real numbers which satisfy the inequality in Corollary \ref{ex_first}. 
\end{kei}
\section{Conclusion}
In this paper, we investigated relationships between formalizations of information-theoretic security for symmetric-key encryption and key-agreement protocols in a general setting. Specifically, we showed that, for symmetric-key encryption, the following formalizations are all equivalent without any condition on system parameters: 
\begin{itemize}
\item Stand-alone security including formalizations of extended (or relaxed) Shannon's secrecy using the statistical distance, information-theoretic indistinguishability and semantic security by Goldwasser and Micali; and  
\item Composable security including formalizations of Maurer et al. and Canetti. 
\end{itemize}  
In addition, we have shown that there are two security formalizations which are not equivalent to the above formalizations without a certain condition: one is the formalization of extended (or relaxed) Shannon's secrecy using the mutual information, and the other is the formalization given by the difference between the distribution of plaintexts and the one conditioned on a certain ciphertext. However, these two formalizations will be equivalent to others, if we impose a certain condition which seems to be satisfied in a usual designing of protocols.  

Furthermore, we also derived lower bounds on the adversary's (or distinguisher's) advantage and secret-key size required under all of the above formalizations. In particular, we could derive them all at once through our relationships between the formalizations. In addition, we briefly observed impossibility results which easily follow from the lower bounds.  

Moreover, we showed similar results (i.e., relationships between formalizations of stand-alone and composable security, lower bounds, and impossibility results) for key agreement protocols. 

We hope that our results on relationships between security formalizations shown by a formal and rigorous way are useful in designing the protocols by selecting suitable system parameters. In particular, our results explicitly imply that encryption and key agreement protocols defined by stand-alone security remain to be secure even if they are composed with other ones, though it may be implicitly assumed by some researchers that the stand-alone security formalizations are sufficient for providing composable security in the information-theoretic settings.

\bigskip

\noindent
{\bf Acknowledgments.} 
The first and second authors would like to thank Hideki Imai, Ryutaroh Matsumoto, and Yusuke Sakai for their helpful comments on the earlier version of this paper.  
The third author would like to thank Ueli Maurer for introducing and explaining him the framework of constructive cryptography and its related topics, when he was visiting ETH Z\"{u}rich, Switzerland.  
Finally, the authors would like to thank anonymous referees for helpful comments on the earlier versions of this paper.

\section*{Appendix A: Definition and Inequality} 

\begin{teigi}{\rm 
Let $X$ be a random variable which takes values in a finite set ${\cal X}$. Then, the min-entropy $H_{\infty }(X)$ and the Hartley entropy $H_0(X)$ are defined by 
\begin{eqnarray*}
H_{\infty }(X)= \min_{x\in {\cal X}} \{ - \log P_X(x)\}, \ \ 
H_0(X)= \log \left| \left\{  x\in {\cal X} | P_X(x)>0 \right\} \right|.
\end{eqnarray*}  
}
\end{teigi}

\medskip

\begin{teigi}{\rm 
Let $X$, $Y$, and $Z$ be random variables associated with distributions $P_{X}$, $P_{Y}$, and $P_Z$, respectively. The {\it mutual information between $X$ and $Y$}, denoted by $I(X;Y)$, is defined by 
\begin{eqnarray*}
I(X;Y):=H(X)-H(X|Y), 
\end{eqnarray*} 
where $H(X)$ (resp. $H(X|Y)$) is the Shannon entropy (resp. the conditional Shannon entropy). 
Also, the {\it conditional mutual information of $X$ and $Y$ given $Z$}, denoted by $I(X;Y|Z)$, is defined by 
\begin{eqnarray*}
I(X;Y|Z):=\sum_{z} P_{Z}(z) I(X;Y|Z=z).
\end{eqnarray*} 
}
\end{teigi}

\medskip

\begin{teigi}{\rm 
Let $X$, $Y$, and $Z$ be random variables associated with distributions $P_{X}$, $P_{Y}$, and $P_Z$, respectively, where $X$ and $Y$ take values in a finite set ${\cal X}$. The {\it statistical distance} (a.k.a. variational distance) between two distributions $P_{X}$ and $P_{Y}$, denoted by $\Delta(P_X,P_Y)$, is defined by 
\begin{eqnarray*}
\Delta(P_X,P_Y):=\frac{1}{2} \sum_{x\in {\cal X}}\left| P_{X}(x)-P_{Y}(x) \right|.
\end{eqnarray*}
} 
\end{teigi}

\medskip
 
Also, for conditional probabilities $P_{X|Z}:=P_{XZ}/P_{Z}$ and $P_{Y|Z}:=P_{YZ}/P_{Z}$, the statistical distance between $P_{X|Z}$ and $P_{Y|Z}$, denoted by $\Delta(P_{X|Z},P_{Y|Z})$ (or $\Delta(X,Y|Z)$), can be defined by 
\begin{eqnarray*}
\Delta (P_{X|Z},P_{Y|Z}):=\sum_{z} P_{Z}(z) \Delta (P_{X|Z=z},P_{Y|Z=z}).
\end{eqnarray*} 
Then, by definitions, note that $\Delta (P_{X|Z},P_{Y|Z})=\Delta (P_{ZX},P_{ZY})$. 

In this appendix, for completeness, we describe several inequalities in the following, which are necessary to show the proofs of propositions in this paper. Note that these inequalities are not new.  

\medskip

\begin{meidai}\label{hosoku1}
Let $(X,Y)$ and $(X',Y')$ be random variables associated with two distributions $P_{XY}$ and $P_{X'Y'}$, respectively, in a finite set. Then, we have 
\begin{eqnarray*}
\max\left( \Delta(P_{X},P_{X'}),\Delta (P_{Y},P_{Y'})\right)  \le \Delta(P_{XY},P_{X'Y'}) 
\end{eqnarray*}
\end{meidai}
\begin{IEEEproof}
From the definition of statistical distance, it follows that 
\begin{eqnarray*}  
2\cdot \Delta(P_{XY},P_{X'Y'}) &=& \sum_{x}\sum_{y} \left| P_{XY}(x,y)-P_{X'Y'}(x,y) \right|\\
&\ge& \sum_{x} \left| \sum_{y}P_{XY}(x,y)-\sum_{y}P_{X'Y'}(x,y) \right|\\
&=&\sum_{x} \left| P_{X}(x)-P_{X'}(x) \right|\\
&=&2\cdot \Delta(P_{X},P_{X'}). 
\end{eqnarray*}
\end{IEEEproof}

\medskip

\begin{meidai}\label{hosoku2}
Let $X$ and $X'$ be random variables associated with two distributions $P_{X}$ and $P_{X'}$, respectively, in a finite set. For an arbitrary random variable $Y$ associated with a distribution $P_{Y}$, we have 
$\Delta(P_{XXY},P_{XX'Y})=P(X\not=X')$. 
\end{meidai}
\begin{IEEEproof}
The proof follows from the following direct calculation:  
\begin{eqnarray*}  
2\cdot \Delta(P_{XXY},P_{XX'Y}) 
&=& \sum_{x}\sum_{x'}\sum_{y} \left| P_{XXY}(x,x',y)-P_{XX'Y}(x,x',y) \right|\\
&=& \sum_{x}\sum_{y} \left| P_{XXY}(x,x,y)-P_{XX'Y}(x,x,y) \right| \\
&&\qquad + \sum_{x}\sum_{x'\not=x}\sum_{y} \left| P_{XXY}(x,x',y)-P_{XX'Y}(x,x',y) \right|\\
&=&\sum_{x}\sum_{y}(P_{XY}(x,y)-P_{XX'Y}(x,x,y))+\sum_{x}\sum_{x'\not=x}\sum_{y}P_{XX'Y}(x,x',y)\\
&=&1-\Pr \{X=X'\}+\Pr \{X\not=X'\}\\
&=&2 \Pr \{ X\not=X'\}. 
\end{eqnarray*}
\end{IEEEproof}

\medskip

\begin{kei} \label{kei1} 
Let $X$ and $X'$ be random variables associated with two distributions $P_{X}$ and $P_{X'}$, respectively, in a finite set. Then, we have 
$\Delta(P_{X},P_{X'})\le \Pr \{ X\not=X'\}$. 
\end{kei}
\begin{IEEEproof}
The proof follows from Propositions \ref{hosoku1} and \ref{hosoku2}.
\end{IEEEproof}

\medskip

\begin{meidai}[Pinsker's inequality, Lemma 12.6.1 in \cite{CT}] \label{CT1}
Let $X_1$ and $X_2$ be random variables associated with two distributions $P_{X_1}$ and $P_{X_2}$, respectively, in a finite set. Then, we have 
\begin{eqnarray*}
D(P_{X_1} \parallel  P_{X_2}) \ge \frac{2}{\ln 2} \Delta(P_{X_1}, P_{X_2})^2. 
\end{eqnarray*}
\end{meidai}

\medskip

\begin{kei}\label{CT4}
Let $X$ and $Y$ be random variables associated with two distributions $P_{X}$ and $P_{Y}$, respectively. Then, we have 
\begin{eqnarray*}
I(X;Y) \ge \frac{2}{\ln 2} \Delta(P_{XY}, P_{X}P_Y)^2. 
\end{eqnarray*}
\end{kei}
\begin{IEEEproof}
The proof immediately follows from Proposition \ref{CT1} by setting $P_{X_1}:=P_{XY}$ and $P_{X_2}:=P_{X}P_{Y}$.
\end{IEEEproof}

\medskip

\begin{meidai}[Classical case of Fannes's inequality, Theorem 16.3.2 in \cite{CT}] \label{CT2}
Let $X_1$ and $X_2$ be random variables associated with two distributions $P_{X_1}$ and $P_{X_2}$, respectively, on a finite set ${\cal X}$ such that 
$\Delta(P_{X_1}, P_{X_2}) \le \frac{1}{4}$. 
Then, we have 
\begin{eqnarray*}
\left| H(X_1)-H(X_2) \right| \le - 2 \Delta(P_{X_1},P_{X_2})\log \frac{2\Delta(P_{X_1},P_{X_2})}{|{\cal X}|}.  
\end{eqnarray*}
\end{meidai}

\medskip

\begin{kei} \label{CT3}
Let ${X}$ and ${Y}$ be random variables which take values in finite sets ${\cal X}$ and ${\cal Y}$, respectively. If $\Delta(P_{XY}, P_{X}P_{Y}) \le \frac{1}{4}$, we have   
\begin{eqnarray*}
I(X;Y) \le - 2 \Delta(P_{XY}, P_{X}P_{Y}) \log \frac{2 \Delta(P_{XY}, P_{X}P_{Y}) }{|{\cal X}||{\cal Y}|}.  
\end{eqnarray*}
\end{kei}
\begin{IEEEproof}
The proof immediately follows from Proposition \ref{CT2} by setting $P_{X_1}:=P_{XY}$ and $P_{X_2}:=P_{X}P_{Y}$. 
\end{IEEEproof}

\section*{Appendix B: Technical Propositions}
In this appendix, we show several propositions which are used in this paper.

\medskip

\begin{meidai}\label{hosoku_relation1}
Let $X$ be a random variable which takes values in a finite set ${\cal X}$. 
In addition, let $Y$ and $Z$ be random variables taking values in a finite set ${\cal Y}$ defined by 
\begin{eqnarray*}
Y:=f(X,R), \quad Z:=g(X,R'),
\end{eqnarray*}
where $f,g$ are mappings and $R,R'$ are random variables such that $X,R,R'$ are pairwisely independent.
We define 
\begin{eqnarray*}
\alpha&:=&\sup_{P_X \in {\cal P}({\cal X})}\Delta(P_{XY},P_{XZ}), \\
\beta&:=&\max_{x\in {\cal X}}\Delta(P_{Y|X=x},P_{Z|X=x}), 
\end{eqnarray*} 
where the supremum in $\alpha$ ranges over all distributions $P_X\in {\cal P}({\cal X})$. 
Then, we have $\alpha=\beta$. 
\end{meidai}
\begin{IEEEproof} 
First, we show $\alpha \le \beta$: For an arbitrary distribution $P_X$, we have 
\begin{eqnarray*}
 \Delta (P_{XY},P_{XZ})&=&\frac{1}{2}\sum_{x,y}|P_{X Y}(x,y) - P_{X Z}(x,y) |\\
&= & \frac{1}{2} \sum_{x} P_{X}(x) \sum_{y} |P_{Y|X=x}(y|x)- P_{Z|X=x}(y|x)|\\
&\le& \frac{1}{2} \max_{x} \sum_{y} |P_{Y|X=x}(y|x)- P_{Z|X=x}(y|x)|  \\
&= &  \max_{x} \Delta (P_{Y|X=x}, P_{Z|X=x}).
\end{eqnarray*}
Therefore, we get $\alpha \le \beta$. 

Secondly, we prove $\alpha \ge \beta$: 
Let $x_0\in {\cal X}$ be an element such that it gives $\beta$, i.e., $x_0=\arg \beta$.  
For any $\epsilon>0$, we define a distribution $P_{\hat{X}}$ by 
\begin{eqnarray*}
P_{\hat{X}}(x):=
\left\{
  \begin{array}{l}
    1-\gamma \mbox{  if $x=x_0$},   \\
    \frac{\gamma}{|{\cal X}|-1} \mbox{  otherwise},  \\
  \end{array}
\right.
\end{eqnarray*}
where $\gamma$ is a positive real number such that $\gamma \beta \le \epsilon$. 
Let $\hat{Y}:=f(\hat{X},R)$ and $\hat{Z}:=f(\hat{X},R')$. Then, we have 
\begin{eqnarray*}
\alpha &\ge& \Delta (P_{\hat{X} \hat{Y}},P_{\hat{X} \hat{Z}})\\
&\ge& (1-\gamma) \Delta (P_{\hat{Y}| \hat{X}=x_0},P_{\hat{Z}|\hat{X}=x_0})\\
&=& (1-\gamma) \beta\\
&\ge& \beta-\epsilon.
\end{eqnarray*}
Therefore, we have $\alpha \ge \beta$. 
\end{IEEEproof}

\medskip

\begin{hodai}\label{furoku1}
For a key agreement protocol, we have 
\begin{eqnarray*}
\Pr \{ K_A\not= K_B\} &\le& \Delta(P_{K_A K_B}, P_{KK}) \\
&\le& P(K_A\not= K_B) +\min \{ \Delta(P_{K_A}, P_{K}),\Delta(P_{K_B}, P_{K})\}.   
\end{eqnarray*}
\end{hodai}
\begin{IEEEproof}
Since we can easily see the existence of a distinguisher with advantage $\Pr \{K_A\not= K_B\}$, the first inequality of the two is easy. 
We show the second inequality in the following. 
From triangle inequality, we have 
\begin{eqnarray*}
\Delta(P_{K_A K_B}, P_{KK}) &\le& \Delta(P_{K_A K_B}, P_{K_AK_A})+\Delta(P_{K_AK_A}, P_{KK})\\
&=& \Pr \{K_A\not= K_B\} +\Delta(P_{K_A}, P_{K}). 
\end{eqnarray*}
Similarly, it is shown that $\Delta(P_{K_A K_B}, P_{KK}) \le \Pr \{K_A\not= K_B\} + \Delta(P_{K_B}, P_{K})$. 
\end{IEEEproof}

\medskip

\begin{hodai}\label{lem:binary}
For two binary random variables $X$ and $Y$ over a set $\{0,1\}$, 
and for $\varepsilon \in [0,1]$, the following two inequalities are equivalent:
\begin{eqnarray}
&&\left| \Pr \{ X=Y \} - \sum_{\ell \in \{0,1\}} \Pr \{ X=\ell \} \Pr \{ Y=\ell \} \right| \le \varepsilon, \label{eq:sum}
\\ 
&&\Bigr| \Pr \{ X=Y=\ell \} - \Pr \{ X=\ell \} \Pr \{ Y=\ell \} \Bigl|\, \le \frac{1}{2}\varepsilon 
\quad \mbox{ for every }\ell \in \{0,1\}. \label{eq:resp}
\end{eqnarray}
\end{hodai}
\begin{IEEEproof}
It is sufficient to show that \eqref{eq:sum} $\Rightarrow$ \eqref{eq:resp}, since \eqref{eq:resp} $\Rightarrow$ \eqref{eq:sum} is obvious. 
Letting $P_{XY}$ be a joint probability distribution of $X$ and $Y$ given by TABLE \ref{table:A=2}, 
\eqref{eq:sum} is equivalent to
\begin{eqnarray}\label{eq:abcd}
\bigr|a+d - (a+b)(a+c) - (c+d)(b+d)\bigr|\le \varepsilon. 
\end{eqnarray}
Since it holds that $a+b+c+d=1$, \eqref{eq:abcd} becomes 
$|ad-bc| \le \varepsilon/2$. Furthermore, using $a+b+c+d=1$ again, we have 
\begin{eqnarray*}
\bigl| P_{XY}(0,0) - P_X(0)P_Y(0)\bigr|
= \bigl|a-(a+b)(a+c)\bigr| 
&\le& \frac{\varepsilon}{2}, \\
\bigl| P_{XY}(1,1) - P_X(1)P_Y(1)\bigr|=\bigl|d- (c+d)(b+d)\bigr| 
&\le& \frac{\varepsilon}{2}  
\end{eqnarray*}
which imply \eqref{eq:resp}. 
\end{IEEEproof}

\medskip

\begin{tyuui}
Note that \eqref{eq:sum} $\Rightarrow$ \eqref{eq:resp} does not generally hold if $X$ and $Y$ are not binary random variables.  
\end{tyuui}

\begin{table}[tb]
\caption{$P_{XY}$ and its marginals}
\begin{center}
\begin{tabular}{c|cc|c}
\hline
\hline
$x \backslash y$ &0&1& $P_X(x)$\\
\hline
$0$ &$a$&$b$& $a+b$\\
$1$ &$c$&$d$& $c+d$\\
\hline
$P_Y(y)$ &$a+c$&$b+d$& $1$\\
\hline
\hline
\end{tabular}
\end{center}
\label{table:A=2}
\end{table}
 
\section*{Appendix C: Proof of Lemma \ref{main_ka1}}
Let Supp$(P_{XY})=\{(x,y)| P_{XY}(x,y)>0 \}\subset {\cal X}\times {\cal Y}$ be the support of $P_{XY}$.   
For any $k_A\in {\cal K}$, and $k_B\in {\cal K}$, we define
\begin{eqnarray*}
\Omega_{k_A,k_B}^{\pi,{\cal T}^{\lambda}}:=
\left\{
  \begin{array}{l|l}
    \    & \exists (x,y)\in \mbox{Supp$(P_{XY})$ such that }    \\
    \    & t_i=f_i(x,t_1,\ldots,t_{i-1}) \mbox{ for odd $i$}   \\
    t^{\lambda}=(t_1,t_2,\ldots,t_{\lambda}) \in {\cal T}^{\lambda}    & t_j=f_j(y,t_1,\ldots,t_{j-1}) \mbox{ for even $j$} \\
    \    & k_A=g_A(x,t_1,t_2,\ldots,t_{\lambda})  \\
    \    & k_B=g_B(y,t_1,t_2,\ldots,t_{\lambda})  \\
  \end{array}
\right\}.
\end{eqnarray*} 
For any $(x,y)\in $ Supp$(P_{XY})$, $k_A\in {\cal K}$, and $k_B\in {\cal K}$, we also define
\begin{eqnarray*}
\Omega_{k_A,k_B,x,y}^{\pi,{\cal T}^{\lambda}}:=\left\{
 \begin{array}{l|l}
        \    & t_i=f_i(x,t_1,\ldots,t_{i-1}) \mbox{ for odd $i$}   \\
    t^{\lambda}=(t_1,t_2,\ldots,t_{\lambda}) \in {\cal T}^{\lambda}    & t_j=f_j(y,t_1,\ldots,t_{j-1}) \mbox{ for even $j$} \\
    \    & k_A=g_A(x,t_1,t_2,\ldots,t_{\lambda})  \\
    \    & k_B=g_B(y,t_1,t_2,\ldots,t_{\lambda})  \\
  \end{array}
\right\}.
\end{eqnarray*}

Then, for any simulator $\sigma$, we have 
\begin{eqnarray}
&&\Delta^{\cal D}(\pi ((\LAC \ )^l \| (\RAC )^{l-1} \| \ [\mbox{$P_{XY}$}])),\sigma ([P_K])) \nonumber  \\
&&\ge \frac{1}{2} \sum_{(k_A,k_B,t^{\lambda})\in {\cal K}\times {{\cal K}}\times {\cal T}^{\lambda}} 
\left|   
P_{\pi}(k_A,k_B,t^{\lambda})-P_{\sigma}(k_A,k_B,t^{\lambda})
\right| \nonumber \\
&&=\max_{{\cal B} \subset {\cal K}\times {{\cal K}}\times {\cal T}^{\lambda} } 
\left\{   
P_{\pi}({\cal B})-P_{\sigma}({\cal B})
\right\} \nonumber  \\
&&\ge \sum_{(k_A,k_B), t^{\lambda}\in \Omega_{k_A,k_B}^{\pi,{\cal T}^{\lambda}}} 
\left(   
P_{\pi}(k_A,k_B,t^{\lambda})-P_{\sigma}(k_A,k_B,t^{\lambda})
\right), \nonumber \\
&&= 1- \sum_{(k_A,k_B), t^{\lambda}\in \Omega_{k_A,k_B}^{\pi,{\cal T}^{\lambda}}}  
P_{\sigma}(k_A,k_B,t^{\lambda}), \label{main_ka}
\end{eqnarray}
where $P_{\pi}$ and $P_{\sigma}$ are distributions by the systems $\pi ((\LAC \ )^l \| (\RAC )^{l-1} \| \ [\mbox{$P_{XY}$}])$ and $\sigma ([P_K])$, respectively. 

We now need the following claim. 

\begin{claim}\label{ideal_ka} 
Suppose that $g_A$ and $g_B$ in the key agreement protocol $\pi$ are deterministic. 
Then, we have
\begin{eqnarray*}
\sum_{(k_A,k_B), t^{\lambda}\in \Omega_{k_A,k_B}^{\pi,{\cal T}^{\lambda}}} P_{\sigma}(k_A,k_B,t^{\lambda})
\le  
2^{H_0(X,Y)-H_{\infty }(K)}.
\end{eqnarray*}
\end{claim}
\begin{IEEEproof}
We note that $P_{\sigma}(k_A,k_B,t^{\lambda})=0$ if $k_A\not=k_B$, and that $P_{\sigma}(k_A,k_B,t^{\lambda})=P_{K}(k)P_{\sigma}(t^{\lambda})$ if $k_A=k_B=k\in {\cal K}$. Thus, we have 
\begin{eqnarray}
\sum_{(k_A,k_B), t^{\lambda}\in \Omega_{k_A,k_B}^{\pi,{\cal T}^{\lambda}}} P_{\sigma}(k_A,k_B,t^{\lambda})
&=&\sum_{k} P_{K}(k)  \sum_{t^{\lambda}\in \Omega_{k,k}^{\pi,{\cal T}^{\lambda}}}P_{\sigma}(t^{\lambda}) \nonumber\\
&\le& \frac{1}{2^{H_{\infty }(K)}} \sum_{k} \ \sum_{(x,y)\in {\rm Supp}(P_{XY})}
\ \sum_{t^{\lambda}\in \Omega_{k,k,x,y}^{\pi,{\cal T}^{\lambda}}} P_{\sigma}(t^{\lambda} ) \nonumber \\
&=& \frac{1}{2^{H_{\infty }(K)}}  \sum_{(x,y)\in {\rm Supp}(P_{XY})}
\left( \sum_{k} \sum_{t^{\lambda}\in \Omega_{k,k,x,y}^{\pi,{\cal T}^{\lambda}}} P_{\sigma}(t^{\lambda} ) \right) \nonumber \\
&\le&  \frac{1}{2^{H_{\infty }(K)}}  \sum_{(x,y)\in {\rm Supp}(P_{XY})} 1 \label{siki11}\\
&=& 2^{H_0(X,Y)-H_{\infty }(K)}. \nonumber
\end{eqnarray}
where (\ref{siki11}) follows from $\Omega_{k,k,x,y}^{\pi,{\cal T}^{\lambda}}\cap \Omega_{k',k',x,y}^{\pi,{\cal T}^{\lambda}}=\emptyset $ if $k\not=k'$, since we assume that $g_A$ and $g_B$ are deterministic. 
\end{IEEEproof}

\medskip

We are back to the proof of Lemma \ref{main_ka1}. If $g_A$ and $g_B$ are deterministic, the proof of Lemma \ref{main_ka1} directly follows from (\ref{main_ka}) and Claim \ref{ideal_ka}. 
We next show that the statement of Lemma \ref{main_ka1} is true, even if we remove the assumption.   
Suppose that $g_A$ or $g_B$ is probabilistic. Let ${\cal R}_A$ (resp. ${\cal R}_B$) be a finite set, and suppose that $g_A$ (resp. $g_B$) chooses a random number $r_A\in {\cal R}_A$ (resp. $r_B\in {\cal R}_B$) according to a probability distribution $P_{R_A}$ (resp. $P_{R_B}$). For each fixed $(r_A,r_B)\in {\cal R}_A \times {\cal R}_B$, a key agreement protocol $\pi_{(r_A,r_B)}$ is specified in which $g_A$ with inputting $r_A$ and $g_B$ with inputting $r_B$ are deterministic. Therefore, we can apply the lower bound derived before. Hence, even if $g_A$ (resp. $g_B$) chooses $r_A\in {\cal R}_A$ (resp. $r_B\in {\cal R}_B$) according to $P_{R_A}$ (resp. $P_{R_B}$), this lower bound cannot be improved. Therefore, the proof of the lemma is completed.

\end{document}